\documentclass[twocolumn,preprintnumbers]{revtex4}
\usepackage{amsmath}
\usepackage{dcolumn}
\usepackage{bm}
\usepackage{epsfig}
\usepackage{epstopdf}
\usepackage{graphicx}
\usepackage{amsfonts}
\usepackage{amssymb}
\usepackage{mathrsfs}
\usepackage{color}
\usepackage{multirow}
\usepackage{appendix}

\begin{document}
\title{Single-Atom Verification of the Information-Theoretical Bound of Irreversibility at the Quantum Level}
\author{J. W. Zhang$^{1,3}$}
\author{K. Rehan$^{1,3}$}
\author{M. Li$^{2}$}
\author{J. C. Li$^{1,3}$}
\author{L. Chen$^{1}$}
\author{S.-L. Su$^{2}$}
\author{L.-L. Yan$^{2}$}
\email{llyan@zzu.edu.cn}
\author{F. Zhou$^{1}$}
\email{zhoufei@wipm.ac.cn}
\author{M. Feng$^{1,2,4}$}
\email{mangfeng@wipm.ac.cn}
\affiliation{$^{1}$ State Key Laboratory of Magnetic Resonance and Atomic and Molecular Physics,
Wuhan Institute of Physics and Mathematics, Innovation Academy of Precision Measurement Science and Technology, Chinese Academy of Sciences, Wuhan, 430071, China\\
$^{2}$ School of Physics, Zhengzhou University, Zhengzhou 450001, China \\
$^{3}$ School of Physics, University of the Chinese Academy of Sciences, Beijing 100049, China\\
$^{4}$ Department of Physics, Zhejiang Normal University, Jinhua 321004, China }

\begin{abstract}
Quantitative measure of disorder or randomness based on the entropy production characterizes thermodynamical irreversibility, which is relevant to the conventional second law of thermodynamics. Here we report, in a quantum mechanical fashion, the first theoretical prediction and experimental exploration of an information-theoretical bound on the entropy production. Our theoretical model consists of a simplest two-level dissipative system driven by a purely classical field, and under the Markovian dissipation, we find that such an information-theoretical bound, not fully validating quantum relaxation processes, strongly depends on the drive-to-decay ratio and the initial state. Furthermore, we carry out experimental verification of this information-theoretical bound by means of a single spin embedded in an ultracold trapped $^{40}$Ca$^{+}$ ion. Our finding, based on a two-level model, is fundamental to any quantum thermodynamical process and indicates much difference and complexity in quantum thermodynamics with respect to the conventionally classical counterpart.
\end{abstract}
\maketitle

\section{INTRODUCTION}
Unitary operations, demonstrating reversibility, are the key concept in every textbook of quantum mechanics. However, perfect isolation of quantum system is not practically possible due to the fact that any realistic system is subject to the coupling to an uncontrollable reservoir. So quantum systems must be regarded as open systems, in which irreversibility of state evolution is overwhelming. On the other hand, the classical concept of irreversibility is mainly associated with the second law of thermodynamics (SLT), arguing that in an isolated system the thermodynamic entropy, behaving as arrow of time, never decreases.

Over the past two decades, the nonequilibrium processes in thermodynamics have drawn much attention since the conventional equilibrium thermodynamics
cannot reasonably treat most natural or engineered processes that occur far from equilibrium. One of the typical approaches for treating nonequilibrium processes is the fluctuation theorem \cite{flu1,flu2}, which compares the forward process to its time reverse, connecting the fundamental time-reversal symmetry of the underlying microscopic dynamics to the thermodynamics.
For example, the fluctuation theorem \cite{crook} has reproduced the Jarzyski equality \cite{jar}, the only equality so far in nonequilibrium thermodynamics, based on the assumption of microscopically reversible and thermostated dynamics. In addition, the fluctuation theorem has also derived the result of nonnegativity of the entropy production, i.e., the essence of the SLT.
Based on this idea, some recent publications \cite{non1,non2,non3,non4,non5} have tried to develop theories to further understand the thermodynamic irreversibility inherent to nonequilibrium processes. In particular, a very recent work focusing on thermal relaxation processes acquired an information-theoretical bound of irreversibility \cite{prl-123-110603}, which imposes a stronger constraint on the entropy production than the conventional SLT.

In addition to the study of nonequilibrium thermodynamics, there has been a parallel line of development exploring thermodynamics on the microscopic scale \cite{QT1,QT2,QT3,quantumther}. Results obtained so far have shown that conventional thermodynamics, describing how large numbers of particles behave, operates differently in quantum regime. Thus the thermodynamic quantities need to be redefined (or re-understood) and the conventional SLT should be modified \cite{QT4,QT5,QT7,bending}. For example, from the quantum perspective, the origin of fluctuations is not just thermal but also quantum, and the von Neumann entropy is irrelevant to the thermodynamic arrow of time, but the characteristic of the state \cite{vedral}.

In the present paper, we explore, both theoretically and experimentally, the information-theoretical bound of irreversibility studied classical mechanically in \cite{prl-123-110603}, in quantum regime. By introducing quantum information concepts, we first deliberate on this bound of irreversibility, under different conditions, in a dissipative process including both the thermal relaxation and decoherence. Amazingly, counter-intuitive scenarios are found, for some values of drive-to-decay ratio (DDR) and initial states of the system, that the bound can be violated. Particularly, by precisely manipulating a single ultracold trapped $^{40}$Ca$^{+}$ ion, we demonstrate experimentally a single-spin verification of this bound, witnessing the predicted violation.

\section{Theory}
We first review briefly the main points in \cite{prl-123-110603}. A typical thermal relaxation process is treated by the variational principle using two quantities, i.e., the entropy production from thermodynamics and the relative entropy (also called Kulback-Leibler divergence) from information theory. Considering variation of the entropy production $\sigma$ in a duration from $\tau=0$ to $\tau=t$, one may obtain the following inequality,
\begin{equation}
\sigma_{[0,t]}\geq D(\rho_{sys}(0)\Vert\rho_{sys}(t)), \label{Eq1}
\end{equation}
where $D(\rho_{sys}(0)\Vert\rho_{sys}(t))$ is the relative entropy representing the Kulback-Leibler divergence of the system's state $\rho_{sys}(t)$ from the state $\rho_{sys}(0)$. As this divergence is non-negative, Eq. (\ref{Eq1}) imposes a more strict bound than the conventional SLT with the bound of zero.

Compared to classical dissipation, a quantum dissipative process involves decoherence in addition to the thermal relaxation. However, following the idea in \cite{prl-123-110603} that focuses on
a dissipative process with a time-independent Hamiltonian of the system, we find that, Eq. (\ref{Eq1}) still validates the variation of the entropy production in
quantum dissipative processes under the condition of no system-bath correlation. As such, we may redefine the entropy production and relative entropy in quantum way, i.e., by von Neumann entropy. The entropy production $\sigma_{[0,t]}=\int_0^{t}d\tau\sigma(\tau)$, with definition of $\sigma(\tau)=-\text{Tr}[\rho(\tau)\ln\rho(\tau)]$, and the relative entropy is defined as $D(\rho_1\Vert\rho_2)=\text{Tr}[\rho_1\ln(\rho_1/\rho_2)]$. Considering a classical bath with much bigger size than the system, we focus our study on the system, which has no work change due to the time-independent Hamiltonian and has no correlation with the bath. So we have the total entropy $\sigma(t)=\sigma_{sys}(t)+\sigma_{bath}(t)$ \cite{lutz}, where the entropies of the system and bath reflect, respectively, the heat change in the system and the heat flow into the bath.
Their von Neumann entropies are given by $\sigma_{sys}=-\text{Tr}[\rho_{sys}\ln\rho_{sys}]$ and $\sigma_{bath}=-\beta\sum_i E_i\lambda_i$, respectively, in which $\beta$ is the usually defined inverse temperature, and the parameters $\lambda_i$ and $E_i$ correspond to the $i$th eigenvalue and the energy of $i$th eigenstate of $\rho_{sys}$.
To reveal whether  Eq. (\ref{Eq1}) really holds or not, we introduce a balance parameter $\Upsilon(t)=\sigma_{[0,t]}- D(\rho_{sys}(0)\Vert\rho_{sys}(t))$ for estimate.

\section{The scheme and the system}

\begin{figure}[hbtp]
\centering {\includegraphics[width=8.8 cm, height=5.5 cm]{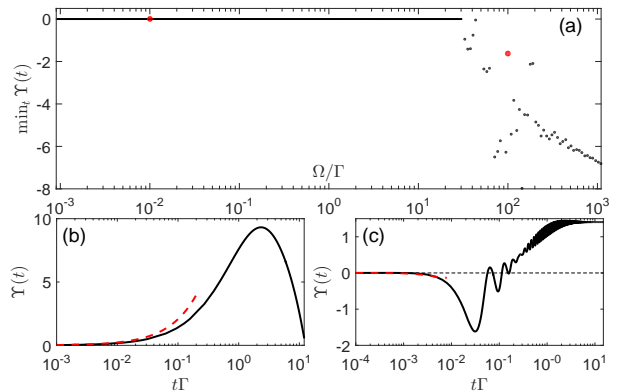}}
\caption{Balance parameter $\Upsilon(t)$ in a dissipative two-level system initialized from the upper level. (a) Minimum values of the balance parameter $\Upsilon(t)$ with respect to the drive-to-decay ratio (DDR), where $\Omega$ is the Rabi frequency regarding the drive and $\Gamma$ is the decay rate. The two red dots indicate, respectively, the scenarios studied in panels (b) and (c). (b) Small DDR case with $\Omega/\Gamma=10^{-2}$ and (c) ultra-large DDR case with $\Omega/\Gamma=10^{2}$. The black solid curves indicate the numerical results of time evolutions of $\Upsilon(t)$, and  the red dashed curves fitted in short time are the analytical results in the cases of $\Omega/\Gamma\ll$1 and $\Omega/\Gamma\gg$1.}
\label{Fig1}
\end{figure}

For the sake of generality, we consider a driven two-level system, which is the fundamental element of any quantum system.
The Hamiltonian in interaction representation is given by $H=\Omega\sigma_{x}/2$, where $\Omega$ is the Rabi frequency, $\sigma_x= |2\rangle\langle 1| + |1\rangle\langle 2|$ is the usual Pauli operator with the upper (lower) level $|2\rangle$ ($|1\rangle$), and we assume $\hbar=1$. To understand the quantum relaxation process, we solve the Lindblad master equation of $H$ as $ \dot{\rho}=-i[H,\rho]+\Gamma(2\sigma_-\rho\sigma_+-\sigma_+\sigma_-\rho-\rho\sigma_+\sigma_-)/2 $ with $\Gamma$ the decay rate from the excited state to the ground state. When $t\rightarrow\infty$, the system reaches the steady state whose eigenspectrum decomposes as $\rho_s=\lambda_-|\phi\rangle_{-}\langle\phi|+\lambda_+|\phi\rangle_{+}\langle\phi|$ with $\lambda_->\lambda_+$ (See Appendix A).  In the subspace spanned by $|\phi\rangle_{-}$ and $|\phi\rangle_{+}$, we find that $|\phi\rangle_{+}$ is of the higher energy than $|\phi\rangle_{-}$, and the effective inverse temperature of the steady state is given by $\beta=(\ln\lambda_--\ln\lambda_+)/\Delta E$ with $\Delta E$ denoting the energy difference between $|\phi\rangle_{+}$ and $|\phi\rangle_{-}$. Thus the corresponding Gibbs state is $\rho_g=e^{-\beta H_i}/Z_g$, where $Z_g$ is the partition function and the effective Hamiltonian $H_i= \Delta E\sigma^{int}_z/2$ with $\sigma^{int}_z=|\phi\rangle_{+}\langle\phi|-|\phi\rangle_{-}\langle\phi|$. A nonequilibrium state of the system evolves to the Gibbs state $\rho_g$ under a quantum relaxation process, implying $\rho_s=\rho_g$ at the end. In this case, we obtain the partition function $Z_g=1/\sqrt{\lambda_-\lambda_+}$.

In terms of the restricted conditions as listed in Appendix A, we segment the problem into four regimes based on the DDR values: small (i.e., $\Omega/\Gamma < 0.15$), intermediate (i.e., $0.15 \leq\Omega/\Gamma < 6.25$), large (i.e., $6.25 \leq\Omega/\Gamma < 25$) and ultra-large (i.e., $\Omega/\Gamma\geq 25$). Since the regime is segmented by $\Omega/\Gamma$, instead of $\Omega$ or $\Gamma$ alone, the values of $\Omega$ and $\Gamma$ can be very small even in the ultra-large DDR case. This implies that the Lindblad master equation under Markovian approximation validates all these segments \cite{breuer}. For a comparison with the classical situation, we first consider, both theoretically and experimentally, the system initialized from a well-polarized state, i.e., $\rho_i=|2\rangle\langle 2|$. For such a non-equilibrium process undergoing a quantum relaxation until the Gibbs state $\rho_g$, Fig. \ref{Fig1}(a) presents a numerical check of the balance parameter over the whole regime, where two extreme cases can be treated analytically as below. \\
{\noindent (I) $\Omega/\Gamma\ll$1, with the time unit $\tau_a\approx\Gamma^{-1}$. For $t\ll\tau_a$, we obtain $\beta=4(\ln\Gamma-\ln\Omega)/\Delta\tilde{E}$ with the energy difference $\Delta\tilde{E}$ between $|2\rangle$ and $|1\rangle$. Thus the balance parameter $\Upsilon(t)$ in the short-time relaxation process, i.e., $t\ll\tau_a$, is
$\Upsilon(t)=H(\Gamma t)+4\Gamma t\ln (\Gamma/\Omega)-\Gamma t$,
where $H(\circ)$ denotes the binary entropy of $\circ$. Due to $H(\Gamma t)>\Gamma t$ for $\Gamma t\ll 1$, we have $\Upsilon(t)>4\Gamma t$, implying that Eq. (\ref{Eq1}) fully holds in the very small DDR case, as plotted in Fig. \ref{Fig1}(b).

{\noindent (II) $\Omega/\Gamma\gg$1, with the time unit $\tau_c\approx\Omega^{-1}$. For $t\ll\tau_c$, the inverse temperature is written as $\beta=4/\Delta\tilde{E}$, independent of $\Gamma$ and $\Omega$. As a result, in the preliminary region of the evolution, $\Upsilon(t)$ is given by $\Upsilon(t)=[2\Gamma-\Omega(\ln 4-1-2\ln \Omega t)]\Omega t^{2}/4$,
which turns to be $\Upsilon(t)\sim -\Omega^2 t^2 <$0 in the short time limit. This implies violation of the information-theoretical bound of irreversibility in this ultra-large DDR case. However, the situation is more complicated than a simple occurrence of violation. If we observe the evolution for a time longer than $t_c$ with $t_c=2e^{-(\Omega+2\Gamma)/2\Omega}\tau_c$, we may find the revival of the information-theoretical bound, see Fig. \ref{Fig1}(c). The violation and validity of the bound could happen repeatedly, which is resulted from the competition between coherence and relaxation regarding the system.

For the cases sandwiched by the two extreme cases, we have to investigate numerically, along with the experimental execution as elucidated below. Our experiment is carried out on a single ultracold $^{40}$Ca$^{+}$ ion confined in a linear Paul trap as employed previously \cite{SA-2-e1600578,njp-19-063032}. Under the pseudo-potential approximation, the axial and radial frequencies of the trap potential are, respectively, $\omega_z/2\pi=1.0$ MHz and $\omega_r/2\pi=1.2$ MHz. For our purpose, we employ a magnetic field of 0.6 mT directed in axial orientation, yielding the ground state $4^2S_{1/2}$ and the metastable state $3^2D_{5/2}$ split into two and six hyperfine energy levels, respectively. We encode qubit $\mid\downarrow\rangle$ in $|4^{2}S_{1/2}, m_{J}=+1/2\rangle$ and $\mid\uparrow\rangle$ in $|3^{2}D_{5/2}, m_{J}=+5/2\rangle$, where $m_{J}$ represents the magnetic quantum number, and for simplicity the two levels are labeled as $|1\rangle$ and $|2\rangle$ (See Fig. \ref{Fig2}).
Before our implementation, the $z$-axis motional mode of the ion is cooled down to average phonon number of $0.030(7)$. The qubit is manipulated by a narrow-linewidth Ti:sapphire laser with wavelength around 729 nm, which irradiates the ultracold ion under the government of the carrier-transition Hamiltonian $H_c=\Omega(\sigma_+e^{i\phi_{L}}+\sigma_-e^{-i\phi_{L}})/2$, with the Rabi frequency $\Omega$ as the laser-ion coupling strength in units of $\hbar=1$. $\phi_{L}$ is the laser phase and $\sigma_{+,-}$ are Pauli operators for the qubit levels.

\begin{figure}[hbtp]
\centering {\includegraphics[width=3.3 cm, height=4.0 cm]{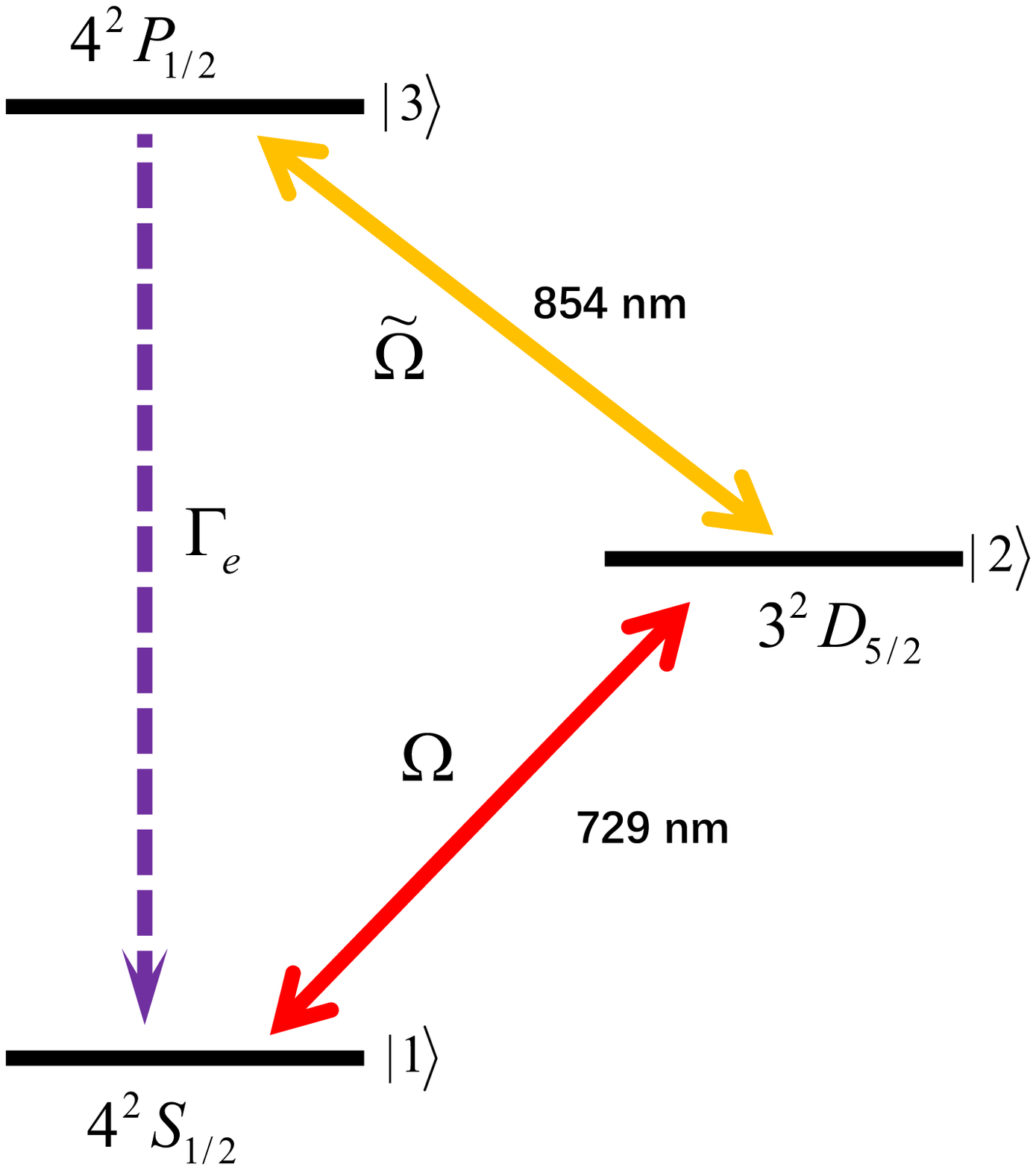}}\centering{\includegraphics[width=5.5 cm, height=4.0 cm]{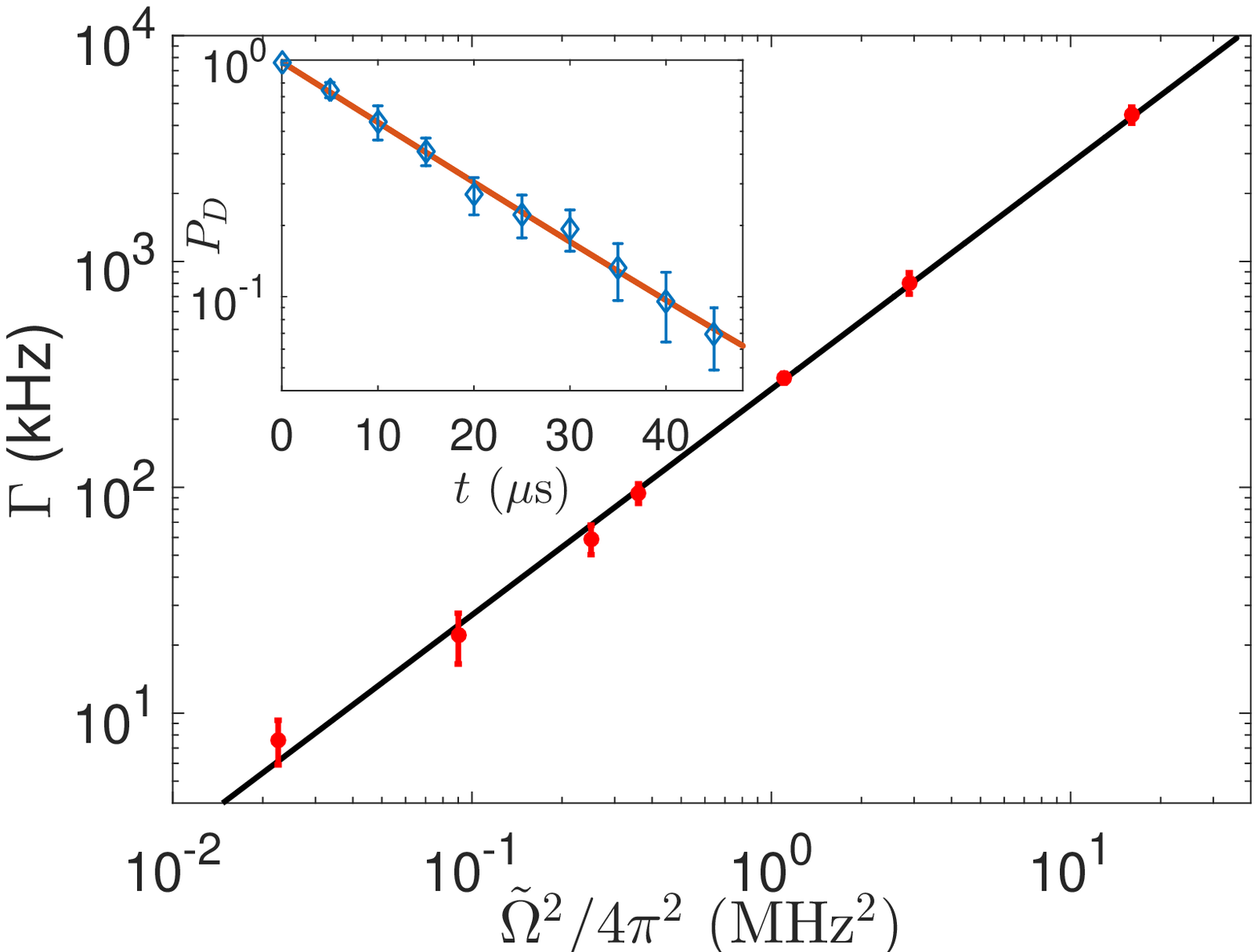}}
\caption{(Left) Level scheme of the $^{40}$Ca$^+$ ion confined in a linear Paul trap system, where only the Zeeman sublevels are plotted for a closed cycle required by an effective two-level system. We label $|1\rangle$, $|2\rangle$ and $|3\rangle$ for the ground state, the metastable state and the excite state, respectively. The lifetimes of the excited state and the metastable state are $6.9$ ns and $1.2$ s, respectively. The wavelengths and the Rabi frequencies of the coupling lasers are indicated, and the decay rate of $|3\rangle$ is $\Gamma_e/2\pi=23.1$ MHz.
(Right) Effective decay rate $\Gamma$ of $|2\rangle$ as a function of the coupling strength $\tilde{\Omega}$ in the case of $\Omega=0$, where dots are experimental measurements and the line is a fitting by the analytical form $\Gamma=\tilde{\Omega}^2/\Gamma_e$ in the case of $\Gamma_e\gg\tilde{\Omega}$. Inset: Time evolution of the population in $|2\rangle$, labeled as $P_{D}$, from which the effective decay rate from $|2\rangle$ to $|1\rangle$ is evaluated as $\Gamma=59(9)$ kHz, in the case of $\tilde{\Omega}/2\pi=500$ kHz. The error bars indicate standard deviation containing the statistical errors of 10,000 measurements for each data point.}
\label{Fig2}
\end{figure}

The single ultracold trapped ion is an ideal platform to explore the thermodynamics due to flexible modeling and ultimate accuracy \cite{QTE1,QTE3,QTE4,QTE5,QTE6}. Here we intend to manifest a dissipative two-level system, and thus we employ additionally the excited level $|4^{2}P_{3/2}, m_{J}=+3/2\rangle$ labeled as $|3\rangle$, with which we have a closed
cycle $|1\rangle$ $\rightarrow$ $|2\rangle$ $\rightarrow$ $|3\rangle$ $\rightarrow$ $|1\rangle$, as presented in Fig. \ref{Fig2}. The first step from $|1\rangle$ to $|2\rangle$ is achieved by the Ti:sapphire laser (729-nm) tuned exactly to the resonance transition. The second step from $|2\rangle$ to $|3\rangle$ is a dipolar transition made by a semiconductor laser (854-nm) under the restriction of the selection rule. The third step $|3\rangle$ $\rightarrow$ $|1\rangle$ is a spontaneous emission, also restricted by the selection rule. Practically, the Rabi frequency $\Omega$ and the dissipative rate $\Gamma$ of the effective two-level system are tuned by tuning the 729-nm and 854-nm lasers under the condition of $\Omega\ll\tilde{\Omega}$. This makes sure that we may check the dissipation each time, as exemplified in the right panel of Fig. \ref{Fig2}, before making measurements for verifying Eq. (\ref{Eq1}). The observed linear decay of $P_{D}$ validates the Lindblad master equation in treating all the cases of the DDR as elucidated below.

\section{Experimental observation}
Our observations are based on different values of DDR and different initial states of the system. So we divide this section into two parts relevant to, respectively, the system initialized from a well-polarized state and from states with coherence. In each part, observations due to some typical DDR values are demonstrated and analyzed.

\begin{figure*}[hbtp]
\centering {\includegraphics[width=19 cm, height=8.2 cm]{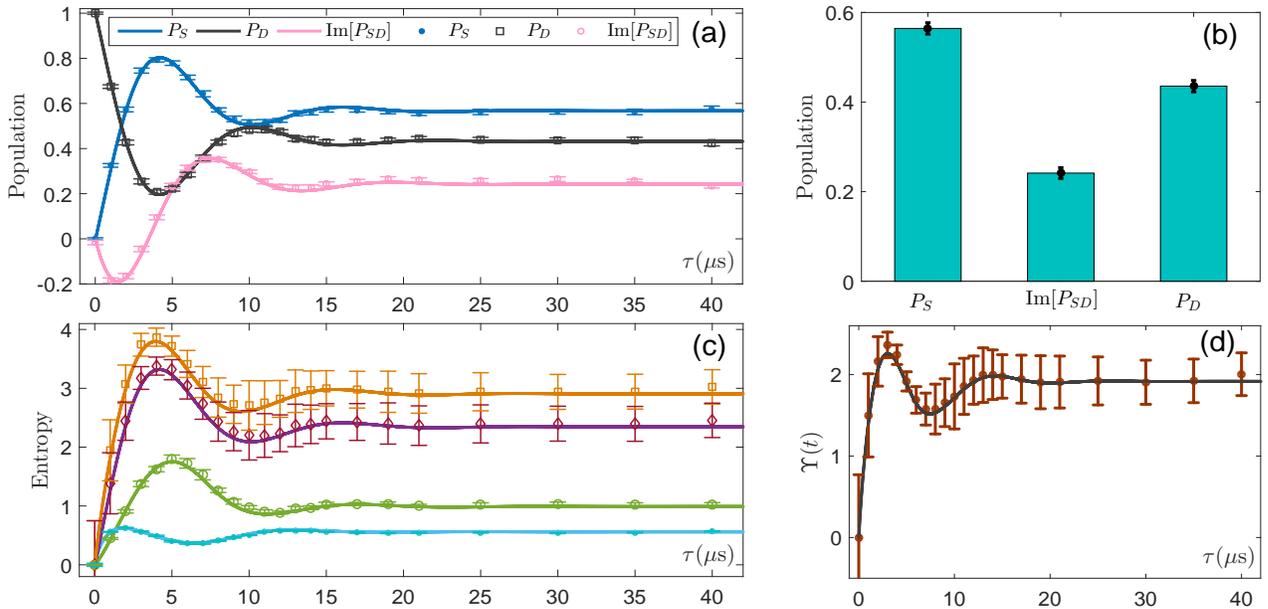}}
\caption{Intermediate DDR with $\Omega/\Gamma=1.8$ and the initial state $|2\rangle$.
(a) Population regarding the state elements in time evolution, where the blue and black curves denote the diagonal terms of the density matrix, and the pink curve corresponds to the off-diagonal term of the density matrix. $P_{S}$ is the population in level $|1\rangle$. (b) Population of the steady state observed at $\tau=50$ $\mu$s. (c) Time evolution of the characteristic entropies, where the curves from the bottom to top (according to the order after $\tau=5$ $\mu$s) represent, respectively, the system's entropy production regarding $\sigma_{sys}$, the relative entropy $D(\rho_{sys}(0)\Vert\rho_{sys}(\tau))$, the bath's entropy production regarding $\sigma_{bath}$, and the total entropy production $\sigma_{[0,t]}$. (d) Time evolution of the balance parameter. The curves and dots in each panels indicate the numerical and experimental results, respectively, where $\Gamma=300$ kHz, $\Omega/2\pi=85$ kHz, $\tilde{\Omega}/2\pi=1$ MHz and $\Gamma_e/2\pi=23.1$ MHz.  The error bars are standard deviation including the statistical errors of 10,000 measurements for each data point. }
\label{Fig3}
\end{figure*}

\subsection{For initialization from a well-polarized state}
We have experimentally measured three curves, belonging to, respectively, the cases of small, intermediate and large DDRs. Here we first consider the intermediate DDR (Fig. \ref{Fig3}), which displays dynamics under a comparable competition between the drive and dissipation, finally reaching an equilibrium at nearly half of the maximal $P_{D}$, see panels (a,b). Other two panels demonstrate that Eq. (\ref{Eq1}) is always valid in the whole process. Both the Kullback-Leibler divergence and the entropy production are initially zero, and then increase in time evolution with the latter always larger than the former. Despite the robustness of Eq. (\ref{Eq1}), we have observed fluctuations of the entropies in the evolution, along with the population variation, which are very different from the entropies' monotonous increase in the classical counterpart \cite{prl-123-110603}. Moreover, the system's entropy production maximizes at the mixed state with maximal off-diagonal terms, implying farthest away from the pure state. In contrast, the relative entropy maximizes at the state with largest Kullback-Leibler divergence from the initial one (i.e., a well-polarized state), which is relevant to certain occupations in the diagonal and off-diagonal terms of the system's state.
In this context, the larger value of $\Upsilon$ reflects more weight of the mixed state involved in the system, which is in contrast to either the equilibrium described in \cite{prl-123-110603} or the thermal equilibrium in the conventionally thermodynamic perception.

We have also carried out experiments for cases with small and large DDRs initialized from $|2\rangle$, which also validate Eq. (1). As presented in Fig. \ref{Fig4}, the results reflect the weight of mixed states, as represented by $\Upsilon$, in these quantum thermodynamic processes. The different variations of $\Upsilon$ here from in Fig. 3 indicate the facts that (1) the system's entropy production in the small DDR case descent quickly due to dominant dissipation, conforming with the observation of $P_{D}$ fast falling down to nearly zero; (2) The level $|2\rangle$ keeps the population for a relatively long time due to the large DDR, and then decay to $|1\rangle$ in a drastically oscillating fashion, yielding a slow increase of $\Upsilon$ in the former part and then a rich dynamics in the later until a complete equilibration.

In this context, we can understand the breakdown of Eq. (1) under the ultra-large DDR, as shown in Fig. \ref{Fig1}(c), which is due to the system's coherence temporarily dominating over the dissipative effect. Unfortunately, we have no way to reach this regime experimentally for the initial state in the excited state $|2\rangle$. In the present case, violating the bound of irreversibility occurs in the case of $\Omega/\Gamma\ge$ 36 (See Fig. 1), thus observing the invalidity of the bound, under the conditions of $\Gamma_{e}\gg\tilde{\Omega}\gg\Omega\gg\Gamma$, requires very small values of the Rabi frequency and effective decay rate, i.e., $\Omega\le\Gamma_e/36^3$ and $\Gamma\le\Gamma_e/36^4$, which is evidently challenging in our experimental observations. As investigated in Appendices C and D, even if we could meet such stringent conditions, credible measurement is unavailable in this case.

\begin{figure}[hbtp]
\centering {\includegraphics[width=9.5 cm, height=6.0 cm]{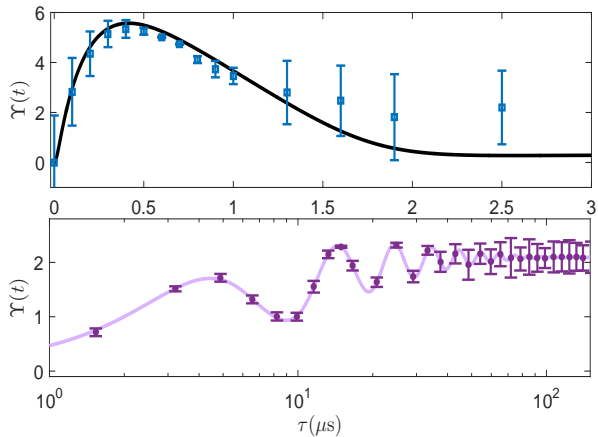}}
\caption{Time evolution of the balance parameter regarding the small DDR (upper panel) and the large DDR (lower panel), where the initial state is $|2\rangle$, and the curves and dots represent numerical results and experimental observations, respectively. In the small DDR situation, $\tilde{\Omega}/2\pi=4$ MHz, $\Omega/2\pi=100$ kHz and $\Gamma=4.5$ MHz, implying $0.09 < \Omega/\Gamma < 0.22$. In the large DDR case, $\tilde{\Omega}/2\pi=470$ kHz, $\Omega/2\pi=100$ kHz and $\Gamma=60$ kHz, meaning $2.5 < \Omega/\Gamma < 12.5$. The error bars are standard deviation covering the statistical errors of 10,000 measurements for each data point. }
\label{Fig4}
\end{figure}

\subsection{For initial states with coherence}
Quantum mechanically, the two-level system could be initialized in a state involving coherence, rather than simply a well-polarized state, for which we present below that Eq. (\ref{Eq1}) is not always valid for most DDR values.

In a quantum two-level system, an arbitrary state can be written as
\begin{equation}
\rho_{i}=\frac{1}{2}(I+\bm{r}\cdot\bm{\sigma})
\end{equation}
where $\bm{\sigma}=(\sigma_x,\sigma_y,\sigma_z)$ and the vector $\bm{r}=(x,y,z)$ with $|\bm{r}|\leq 1$. $|\bm{r}|= 1$ denotes a pure state and $\bm{r}=(0, 0, 1)$ indicates an excited state denoted by $\rho=|2\rangle\langle 2|$. The eigenvalues of $\rho_i$ are written as
\begin{equation}
\bar{\lambda}_{\pm}=\frac{1}{2}(1\pm \vert \bm{r}\vert)  \notag
\end{equation}
correspond to the eigenstates $|\varphi\rangle_{\pm}=a_{\pm}|1\rangle+b_{\pm}|2\rangle$ with
\begin{equation}
|b_{\pm}|^2=\frac{|\bm{r}|\pm z}{2|\bm{r}|}, ~~ a_{\pm}=\frac{\pm |\bm{r}|-z}{x-iy}b_{\pm}.  \notag
\end{equation}
The energies of the two states are given by
\begin{equation}
E_{\pm}=\frac{1}{2|\bm{r}|}[(|\bm{r}|\pm z)E_2+(|\bm{r}|\mp z)E_1],
\end{equation}
with the difference $\Delta E=\frac{z}{|\bm{r}|}(E_2-E_1)$. Since $E_2>E_1$, if $z\geq 0$, then we have $\Delta E\geq 0$ and $|\varphi\rangle_{+}$ is the upper state; if $z\leq 0$, then $\Delta E\leq 0$ and and $|\varphi\rangle_{-}$ is the upper state. The initial entropies of the system and the bath are then given by
\begin{equation}
S_{sys}=-\sum_{k=\pm}\bar{\lambda}_k \ln\bar{\lambda}_{k}, ~~ S_{bath}=-\beta\sum_{k=\pm}E_{\pm}\bar{\lambda}_{\pm}.
\end{equation}

For clarity of description, we employ $\bm{r}$ and $z$ as additional variables, with which the initial state is denoted by the vector $\bm{r}=(x,y,z)$ with $|\bm{r}|\leq 1$. We exemplify the intermediate DDR in Fig. \ref{Fig5}(a) by considering different initial states.
For the initial state with $z<0$, the bound violation would possibly occur, which could be understood as that with the system evolving away from the initially mixed state, the total entropy production, of negative values in some cases, is always smaller than the Kullback-Leibler divergence. Experimentally we have prepared the initial state by evolving for $\tau_0=4$ $\mu s$
from the state $|2\rangle$, and we observed the negative values of $\Upsilon$ in all the evolution, see Fig. \ref{Fig5}(b). By this way, we have found that the invalidity of Eq. (1) could be witnessed more easily with larger values of the DDR.

\begin{figure}[hbtp]
\centering {\includegraphics[width=9.0 cm, height=4.5 cm]{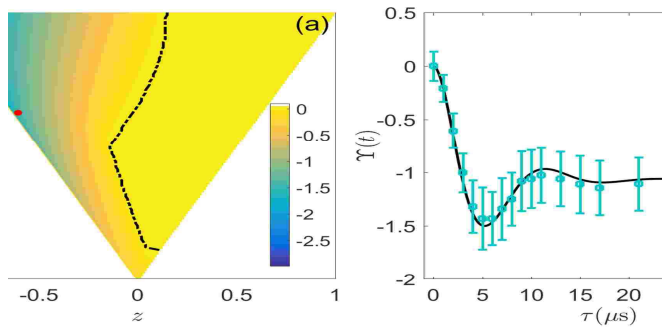}}
\caption{Intermediate DDR with the same values of parameters as in Fig. \ref{Fig3} except the initial states. (a) Minimum values of the balance parameter $\Upsilon(t)$ for different initial states, where $\bm{r}$ is set by $x=0$ and $y=\sqrt{|\bm{r}|^2-z^2}$). The dashed curve denotes the zero values of $\min_t\Upsilon(t)$ and the red dot represents the position of the initial state employed in (b). The color bar indicates the values of $\min_t\Upsilon(t)$. (b) Time evolution of the balance parameter from the initial state denoted by $|\bm{r}|=0.626$ and $z=-0.604$. The curve and dots are the numerical and experimental results, respectively, as in Fig. \ref{Fig3}. }
\label{Fig5}
\end{figure}

\begin{figure}[hbtp]
\centering {\includegraphics[width=9.0 cm, height=4.5 cm]{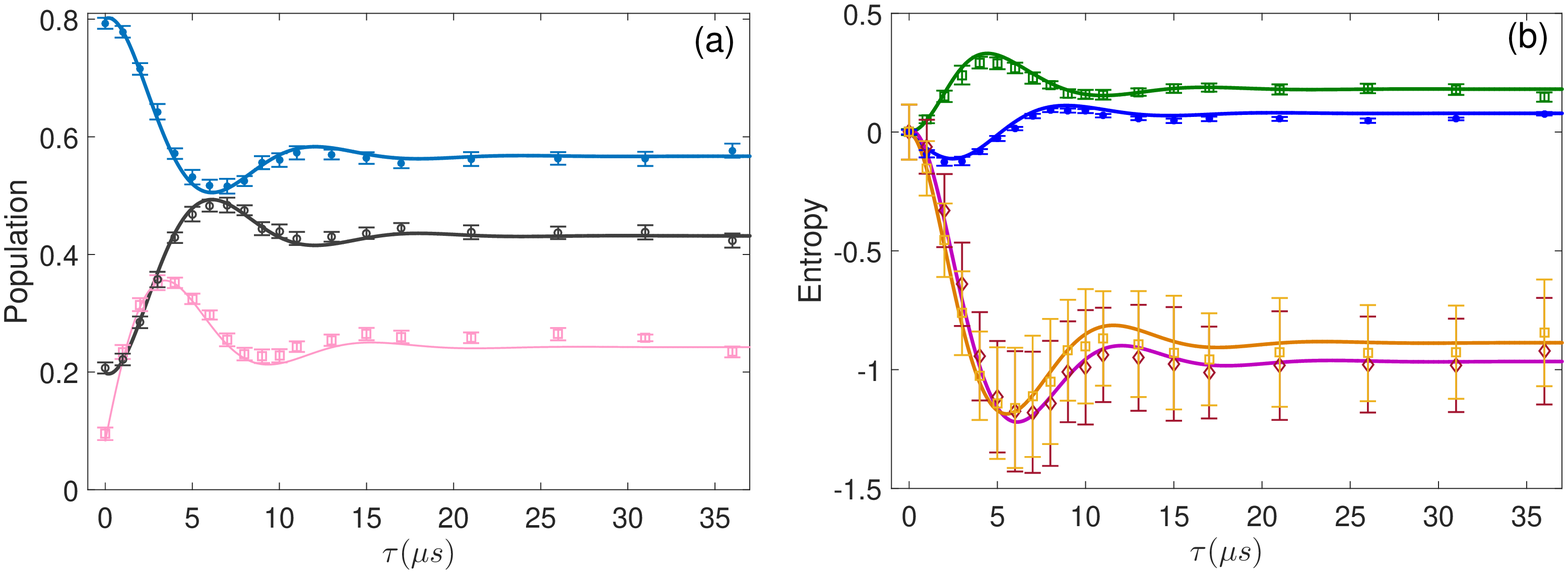}}
\caption{ Supplementary data for Fig. 5. (a) Population regarding the state elements in time evolution, where the blue and black curves denote the diagonal terms $P_S$ and $P_D$ of the density matrix, respectively, and the pink curve corresponds to the off-diagonal term of the density matrix. (b) Time evolution of the characteristic entropies, where the curves from the bottom to top (according to the order after $\tau=10$ $\mu$s) represent, respectively, the bath's entropy production regarding $\sigma_{bath}$, the total entropy production $\sigma_{[0,t]}$, the system's entropy production regarding $\sigma_{sys}$ and the relative entropy $D(\rho_{sys}(0)\Vert\rho_{sys}(\tau))$. The error bars are standard deviation covering the statistical errors of 10,000 measurements for each data point.}
\label{Figs11}
\end{figure}

As a supplement of Fig. 5(b), Fig. \ref{Figs11} presents time evolutions of state populations and characteristic entropies, demonstrating the reason leading to negative values of the balance parameter $\Upsilon(t)$. With the populations oscillating, the total entropy production is always negative but the relative entropy is always positive. This leads to breakdown of the information-theoretical bound of irreversibility in Fig. \ref{Fig5}.

Correspondingly, we may observe the violation of information-theoretical bound of irreversibility in the small and large DDR cases by modifying the initial states in Fig. \ref{Fig5}. As shown in
Fig. \ref{Figs12}, the bound holds for most initial states in the small DDR case, with the violation happening in the limit of $z\rightarrow -|\bm{r}|$. In contrast, in the case of large DDR,
the bound is violated for most initial states and holds only in the limit of $z\rightarrow |\bm{r}|$, as shown in Fig. \ref{Figs13}. However, experimentally, the result in Fig. \ref{Figs12}(b) is not convincing due to the weak violation influenced by the considerably larger errors (which are not resulted from measurement imprecision, but the calculation of entropy, as explained in the next section). In contrast, the experimental result in Fig. \ref{Figs13}(b) clearly shows the violation of the bound.

\begin{figure}[hbtp]
\centering {\includegraphics[width=9.0 cm, height=4.5 cm]{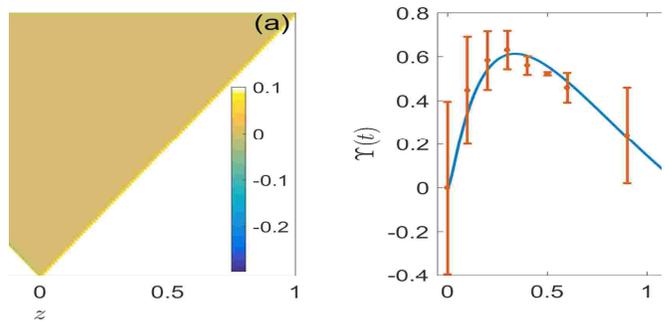}}
\caption{ Small DDR case with the same parameter values as in Fig. \ref{Fig4}(Upper panel) except the initial states. (a) Minimum values of the balance parameter $\Upsilon(t)$ for different initial states, where $\bm{r}$ is set by $x=0$ and $y=\sqrt{|\bm{r}|^2-z^2}$. The dashed curve denotes the zero line of $\min_t\Upsilon(t)$ and the black dot represents the position of the initial state employed in (b). The color bar indicates the values of $\min_t\Upsilon(t)$. (b) Time evolution of the balance parameter from the initial state denoted by $z=-0.654$ and $|\bm{r}|=0.655$. The curves and dots indicate the numerical and experimental results, respectively, where $\Gamma=4.5$ MHz and $\Omega/2\pi=100$ kHz. The error bars are standard deviation covering the statistical errors of 10,000 measurements for each data point. }
\label{Figs12}
\end{figure}

\begin{figure}[hbtp]
\centering{\includegraphics[width=9.0 cm, height=4.5 cm]{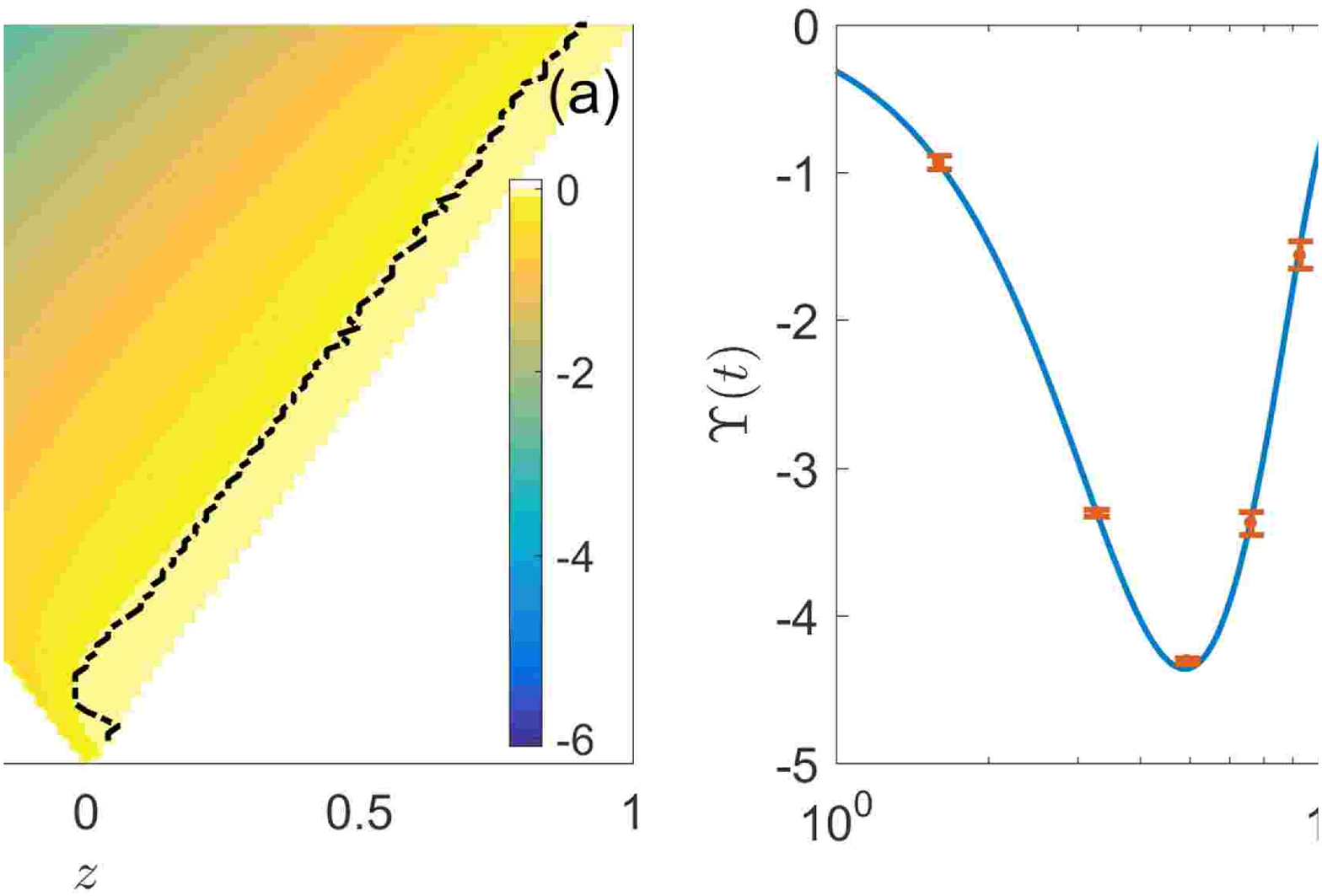}}
\caption{ Large DDR case with the same parameter values as in Fig. \ref{Fig4}(Lower panel) except the initial states. (a) Minimum values of the balance parameter $\Upsilon(t)$ for different initial states, where $\bm{r}$ is set by $x=0$ and $y=\sqrt{|\bm{r}|^2-z^2}$). The dashed curve denotes the zero line of $\min_t\Upsilon(t)$ and the black dot represents the position of the initial state employed in (b). The color bar indicates the values of $\min_t\Upsilon(t)$. (b) Time evolution of the balance parameter from the initial state denoted by $z=-0.811$ and $|\bm{r}|=0.825$. The curves and dots indicate the numerical and experimental results, respectively, where $\Gamma=60$ kHz and $\Omega/2\pi=100$ kHz. The error bars are standard deviation covering the statistical errors of 10,000 measurements for each data point. }
\label{Figs13}
\end{figure}


\section{DISCUSSION}
In our implementation, the decay rate of $|3\rangle$ is $\Gamma_e/2\pi=23.1$ MHz, and the Rabi frequency $\Omega$ can be up to 200 kHz. The effective decay is controlled by the power of the 854-nm laser. As the effective two-level system is operated with population transfer from lower to upper state and decay from the upper to lower state, it is very important to have an exact control of the 729-nm and 854-nm lasers for their polarizations and resonant frequencies. Specifically, since the resonance transition of 729-nm irradiation along x direction is sensitive to the polarization of the laser, we have finely tuned the half-wavelength plate to reach  the maximal resonance strength. Meanwhile, we control the frequency of 854-nm laser by an electric-optic modulator which is locked to another ultra-low expansion cavity (with linewidth of 0.5 MHz). To have an effective decay rate $\Gamma$, we have first swept the frequency of the 854-nm laser to find the exact resonance between $|2\rangle$ and $|3\rangle$, and then determined the expected value of $\Gamma$ by tuning the power of the 854-nm laser.

To make sure our measurements to be performed with high precision, we have tried to suppress quantum projection noise in single-qubit measurements by 10,000 repetition. Moreover,
since the Zeeman sublevel $|3^{2}D_{5/2}, m_{J}=+5/2\rangle$ is highly sensitive to the magnetic field fluctuation, unexpected imperfection is involved in our qubit in initial-state preparations and qubit operations. Other possible errors can also be from the laser instability, imperfect single-qubit pulses and heat noise, whose effects are assessed from the Rabi oscillations in our case. After calibration, we have estimated the total error contributed on the imperfection of the initial-state preparation and the final-state detection to be 0.7(2)$\%$ and 0.22(8)$\%$, respectively. These imperfections, along with statistic errors, are involved in the standard deviation represented by the error bars.

\begin{figure}[hbtp]
\centering {\includegraphics[width=9.2 cm, height=4.8 cm]{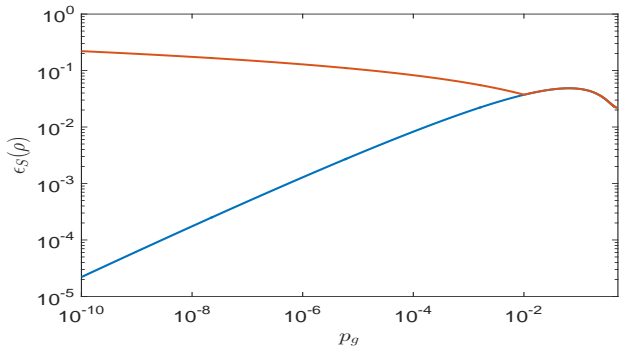}}
\caption{Errors in the calculation of von Neumann entropy for the cases of $p_g\rightarrow 0$, where the upper (lower) curve represents the scenario with (without) the experimental error considered. The parameters are chosen as $N=100$ and $\epsilon_{\text{exp}}=0.01$.}
\label{Figs10}
\end{figure}

Before ending our discussion, we would like to emphasize the intrinsic causes leading to some large error bars regarding our experimental data of entropy. In a general two-level system, the spectrum decomposition of a quantum state $\rho$ can be written as $\rho=p_g|\varphi\rangle_g\langle\varphi|+p_e|\varphi\rangle_e\langle\varphi|$ with $ |\varphi\rangle_{e,g} $ denoting the eigenstates of $\rho$ and $p_e+p_g=1$. In this case, the von Neumann entropy is given by $S(\rho)=-p_g\log p_g-p_e\log p_e$. Theoretically, the error regarding quantum projection noise for detection is estimated as $\epsilon_g=\sqrt{p_g(1-p_g)/N}$ with $N$ denoting the detection repetition. Thus, the error estimate of the von Neumann entropy is formulated as
\begin{eqnarray}
\epsilon_S(\rho)&=&\sqrt{(1+\log p_g)^2\epsilon_g^2+(1+\log p_e)^2\epsilon_e^2} \notag \\
&=&\epsilon_g\sqrt{(1+\log p_g)^2+(1+\log p_e)^2}, \notag
\end{eqnarray}
where we have assumed $\epsilon_g=\epsilon_e$ in the second equation. In the limit $p_g\rightarrow$ 0 and 1, the above equation yields both cases to be $\epsilon_S(\rho)\rightarrow 0$. However, considering the experimental situation, we have, except for the quantum projection error, some other errors as mentioned above, which could not be fully eliminated by the standard calibration method. Therefore, the realistic error in the limit $p_g\rightarrow 0$ is given by,
\begin{equation}
\epsilon_S(\rho)\sim \epsilon_{\text{exp}}\log \frac{1}{p_g}, \notag
\end{equation}
which, as seen in Fig. \ref{Figs10}, is larger than the counterpart with only theoretical consideration. For the case of $p_g\rightarrow 1$, we have similar situation. So we have some of the error bars for entropies evidently larger than the others, e.g., in Figs. 3-8.

\section{CONCLUSION}
We have explored a quantum mechanical information-theoretical bound of irreversibility using a fundamental two-level system, which applies to any quantum thermodynamic process. Our experimental witness is the first single-spin evidence verifying such a novel and tighter bound for irreversibility, which is expected to be helpful in understanding
the SLT and irreversibility subject to quantum effects. Since von Neumann entropy is not relevant to the thermodynamic arrow of time, violating Eq. (1), although counter-intuitive in the viewpoint of classical physics, is possible and reasonable in quantum systems, which means the essential role of coherence as well as the complicated and nontrivial feature in quantum thermodynamics. We believe that our result provides a deeper insight into irreversibility in quantum regime and in particular helps further understanding of thermodynamics at the microscopic scale.


\section*{ACKNOWLEDGEMENTS}
This work was supported by National Key Research $\&$ Development Program of China under grant No. 2017YFA0304503, by National Natural Science Foundation of
China under Grant Nos. 11835011, 11804375, 11804308, 11734018, 11674360 and 61862014. K.R. acknowledges thankfully support from CAS-TWAS president's fellowship. The first two authors JWZ and KR contribute equally in this work.

\appendix

\begin{center}{\bf APPENDIX}\end{center}

We provide below supplementary information to the main text. We first describe the real scenario of simplest two-level system and then present a thorough  transformation of a three-level system to an effective two-level system. The subsequent discussions are based on this effective two-level system. For clarity, we plot Fig. \ref{SFig0} with denotations labeled for convenience of description in Appendices A and B as below. Moreover, the analytical deductions presented below are basically general, but some numerical results exemplified, such as in Figs. 12-15, are made based on the initial state $|2\rangle$. Experimental feasibility for ultra-large DDR case discussed in Appendices C and D is also based on the initial state $|2\rangle$.

\begin{figure}[hbtp]
\centering {\includegraphics[width=9.0 cm, height=4.5 cm]{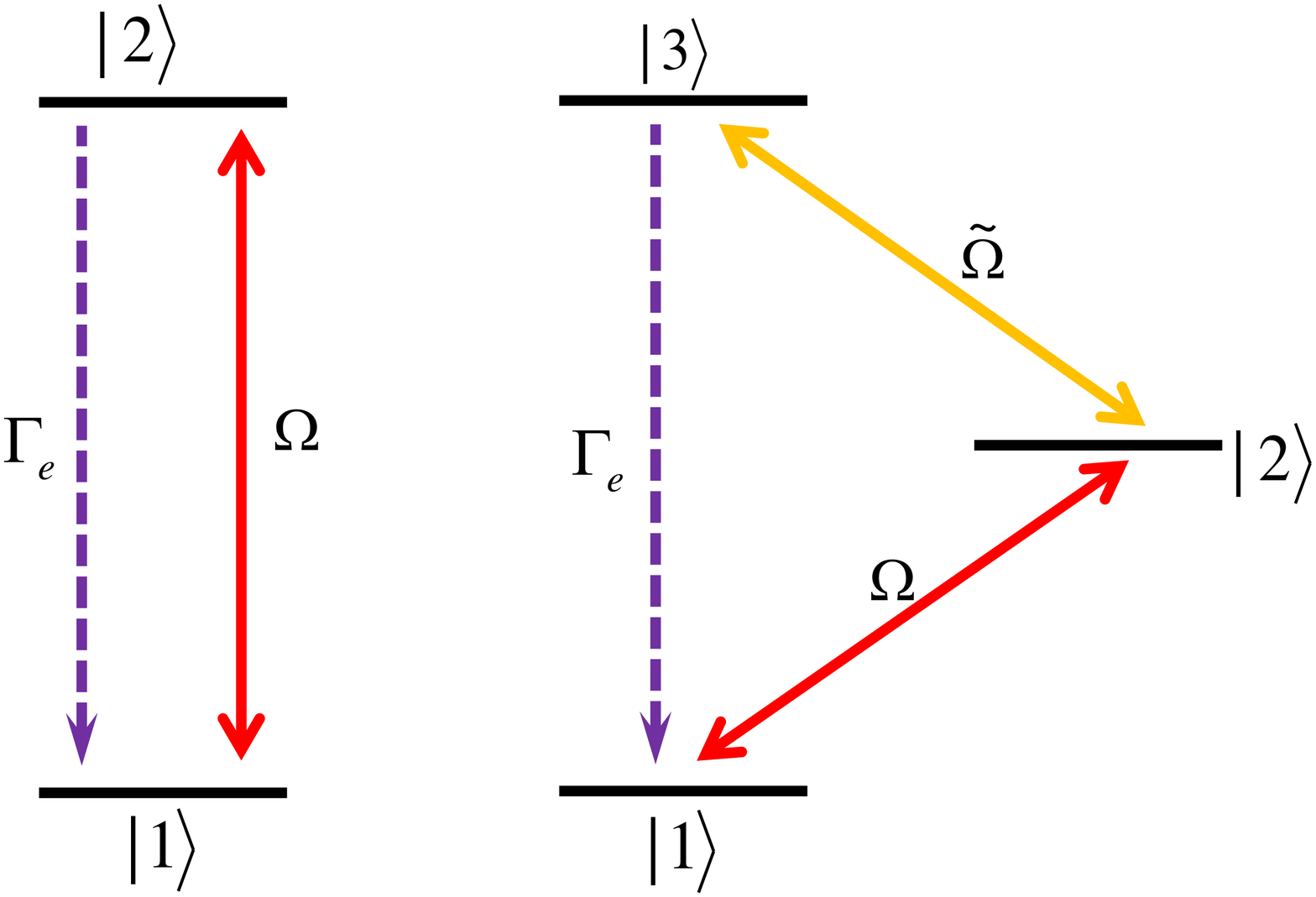}}
\caption{(Left panel) Two-level system discussed in Appendix A, where $\Omega$ and $\Gamma$ are, respectively, the Rabi frequency coupling $|1\rangle$ to $|2\rangle$ and the decay rate from $|2\rangle$. (Right panel) Three-level system discussed in Appendix B, where Rabi frequency $\Omega$ couples $|1\rangle$ to $|2\rangle$, Rabi frequency $\tilde\Omega$ couples $|2\rangle$ to $|3\rangle$, and $\Gamma_{e}$ is the decay rate regarding $|3\rangle$. }
\label{SFig0}
\end{figure}

\section{Quantum relaxation process in a two-level system}
Considering a two-level system such that excited and ground states are, respectively, given as $|2\rangle$ and $|1\rangle$. The mathematical equations arguing the systematic time evolution of the density matrix elements $\rho_{ij}=\langle i|\rho|j\rangle$ can be described as follows,
\begin{eqnarray}
\frac{d}{dt}\rho_{11}&=&-i\frac{\Omega}{2}(\rho_{21}-\rho_{12})+\Gamma\rho_{22}, \notag \\
\frac{d}{dt}\rho_{22}&=&-i\frac{\Omega}{2}(\rho_{12}-\rho_{21})-\Gamma\rho_{22}, \notag \\
\frac{d}{dt}\rho_{12}&=&-i\frac{\Omega}{2}(\rho_{22}-\rho_{11})-\frac{\Gamma}{2}\rho_{12}, \notag \\
\frac{d}{dt}\rho_{21}&=&-i\frac{\Omega}{2}(\rho_{11}-\rho_{22})-\frac{\Gamma}{2}\rho_{21}. \notag
\label{Eqs1}
\end{eqnarray}
Assuming the system initially in $|2\rangle$ and following the aforementioned analytical formulation, we may write,
\begin{eqnarray}
\rho_{22}(t)&=&\frac{1}{2(\Gamma^2+2\Omega^2)\sqrt{\Gamma^2-16\Omega^2}}e^{-\frac{t}{4}(3\Gamma+\sqrt{\Gamma^2-16\Omega^2})} \notag\\
&&[\Gamma(\Gamma^2+5\Omega^2)(1-e^{\frac{t}{2}\sqrt{\Gamma^2-16\Omega^2}})+(\Gamma^2+\Omega^2)\notag \\
&&\sqrt{\Gamma^2-16\Omega^2}(1+e^{\frac{t}{2}\sqrt{\Gamma^2-16\Omega^2}})]+\frac{\Omega^2}{\Gamma^2+2\Omega^2}, \notag\\
\rho_{12}(t)&=&\frac{i\Omega}{2(\Gamma^2+2\Omega^2)\sqrt{\Gamma^2-16\Omega^2}}e^{-\frac{t}{4}(3\Gamma+\sqrt{\Gamma^2-16\Omega^2})}\notag \\
&&[(5\Gamma^2+4\Omega^2)(1-e^{\frac{t}{2}\sqrt{\Gamma^2-16\Omega^2}})-\Gamma\sqrt{\Gamma^2-16\Omega^2} \notag\\
&&(1+e^{\frac{t}{2}\sqrt{\Gamma^2-16\Omega^2}}-2e^{\frac{t}{4}(3\Gamma+\sqrt{\Gamma^2-16\Omega^2})})] \notag.
\end{eqnarray}
For our purpose, we investigate the above dynamic process by taking different time scales into account as below,
\begin{equation}
\tau_a=\frac{4}{3\Gamma+\text{Re}(\sqrt{\Gamma^2-16\Omega^2})}, \quad \tau_b=\frac{4}{3\Gamma-\text{Re}(\sqrt{\Gamma^2-16\Omega^2})} \notag
\end{equation}
and
\begin{equation}
\tau_c=\frac{4}{\text{Im}(\sqrt{\Gamma^2-16\Omega^2})} \notag
\end{equation}
with $\tau_a<\tau_b<2\tau_a$. When $t\ll \min[\tau_a,\tau_c]$ and $\Gamma\sim \Omega$, we obtain
\begin{eqnarray}
\rho_{22}(t)&=&\frac{\Gamma^2+\Omega^2}{\Gamma^2+2\Omega^2}e^{-\frac{t}{4}(3\Gamma+\sqrt{\Gamma^2-16\Omega^2})}+\frac{\Omega^2}{\Gamma^2+2\Omega^2}, \notag\\
\rho_{12}(t)&=&\frac{i\Omega\Gamma}{\Gamma^2+2\Omega^2}[1-e^{-\frac{t}{4}(3\Gamma+\sqrt{\Gamma^2-16\Omega^2})}] \notag.
\end{eqnarray}
It is evident from the above that as the population in $|2\rangle$ descents the coherence terms appears. However, for $t\gg\tau_b$, the system approaches the steady state ($t\rightarrow\infty$), which is given by
\begin{displaymath}
\rho_s=\frac{1}{\Gamma^2+2\Omega^2}
\left( \begin{array}{ccccccccc}
\Gamma^2+\Omega^2 & i\Omega\Gamma\\
-i\Omega\Gamma & \Omega^2
\end{array} \right). \notag
\end{displaymath}
The eigenvalues of the steady state are
\begin{equation}
\lambda_{\pm}=\frac{\Gamma^2+2\Omega^2\mp\Gamma\sqrt{\Gamma^2+4\Omega^2}}{2(\Gamma^2+2\Omega^2)},\notag
\end{equation}
and the corresponding eigenstates are
\begin{equation}
 |\phi\rangle_{+}=(-i\cos\theta_+,\sin\theta_+),\ |\phi\rangle_{-}=(i\cos\theta_-,\sin\theta_-) \notag
\end{equation}
with $\theta_+=\arcsin\frac{2\Omega}{\sqrt{2(\Gamma^2+4\Omega^2-\Gamma\sqrt{\Gamma^2+4\Omega^2})}}$ and $\theta_-=\arcsin\frac{2\Omega}{\sqrt{2(\Gamma^2+4\Omega^2+\Gamma\sqrt{\Gamma^2+4\Omega^2})}}$ satisfying $\theta_+>\theta_-$ and $\theta_-+\theta_+=\pi/2$. Thus, the steady state can be rewritten as
\begin{equation}
\rho_s=\lambda_-|\phi\rangle_{-}\langle\phi|+\lambda_+|\phi\rangle_{+}\langle\phi| \notag
\end{equation}
with $\lambda_->\lambda_+$.

Denoting the energies of $|1\rangle$ and $|2\rangle$ by $E_{1,2}$ with $E_2>E_1$, the corresponding energies of $|\phi\rangle_{\pm}$ are
\begin{equation}
E_{+}=\cos^2\theta_+E_1+\sin^2\theta_+E_2, ~E_{-}=\cos^2\theta_-E_1+\sin^2\theta_-E_2 , \notag
\end{equation}
with their difference given by
\begin{equation}
\Delta E=E_+-E_-=(\sin^2\theta_+-\sin^2\theta_-)(E_2-E_1)>0. \notag
\end{equation}
Therefore, in the picture of $|\phi\rangle_{\pm}$, $|\phi\rangle_{+}$ refers to the excited state and $|\phi\rangle_{-}$ designates the ground state, constituting a new-basis two-level system. In the steady state, the inverse temperature $\beta$ of this new-basis two-level system is given by
\begin{equation}
\beta=\frac{1}{\Delta E}\ln\frac{\lambda_-}{\lambda_+}. \notag
\end{equation}
Since $\lambda_->\lambda_+$, we always have $\beta>0$. From the expressions above, we may rewrite $\beta$ as,
\begin{equation}
\beta=\frac{2}{E_2-E_1}\frac{\sqrt{\Gamma^2+4\Omega^2}}{\Gamma}\ln\frac{\Gamma^2+2\Omega^2+\Gamma\sqrt{\Gamma^2+4\Omega^2}}{2\Omega^2}, \notag
\end{equation}
which implies that different inverse temperature can be obtained by changing the values of two control parameters, i.e, the Rabi frequency $\Omega$ and the decay rate $\Gamma$. In the limit of $\Gamma\gg\Omega$, we obtain
\begin{equation}
\beta=\frac{4}{ (E_2-E_1)}\ln\frac{\Gamma}{\Omega}. \notag
\end{equation}
In contrast, under the condition of $\Gamma\ll\Omega$, the inverse temperature can be expressed as
\begin{equation}
\beta=\frac{4}{(E_2-E_1)}. \notag
\end{equation}
Therefore, for a given decay rate $\Gamma$, the inverse temperature $\beta$ decreases with the increase of $\Omega$, as plotted in Fig. \ref{SFig1}. This also implies that  temperature increases with $\Omega$.

\begin{figure}[hbtp]
\centering {\includegraphics[width=9 cm, height=4.5 cm]{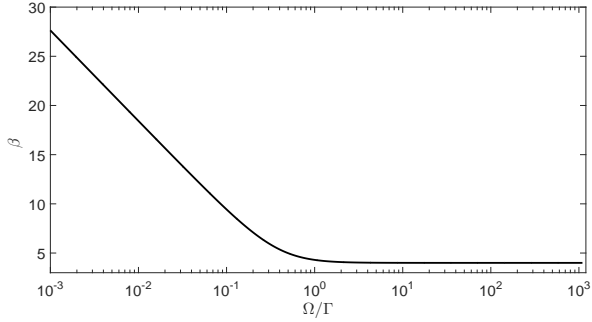}}
\caption{Inverse temperature $\beta$ as a function of $\Omega/\Gamma$ in a two-level system.}
\label{SFig1}
\end{figure}

Now we consider a Gibbs state
\begin{equation}
\rho_g=\frac{1}{Z_g}e^{-\beta H_i},   \notag
\end{equation}
with the partition function $Z_g$ and the Hamiltonian
\begin{equation}
H_i=\frac{1}{2}\hbar \Delta E \sigma^{int}_z    \notag
\end{equation}
in the space spanned by $|\phi\rangle_{-}$ and $|\phi\rangle_{+}$ with $\sigma^{int}_z=|\phi\rangle_{+}\langle\phi|-|\phi\rangle_{-}\langle\phi|$. In the limit of $t\rightarrow\infty$, the quantum relaxation process reaches a steady state as
\begin{equation}
\rho_s=\rho_g.    \notag
\end{equation}
As such, we obtain the partition function $Z_g=1/\sqrt{\lambda_-\lambda_+}$. If we assume the system to be initially in a non-equilibrium state $\rho_{i}$, e.g., $\rho_i=|2\rangle\langle 2|$, then  the ensuing system evolution encapsulating the thermal relaxation process will lead to the Gibbs state $\rho_s$.

In the case of $\Omega/\Gamma\ll$1 with $\tau_a\approx\Gamma^{-1}$, we obtain in the case of $t\ll\tau_a$,
\begin{equation}
\rho_{22}(t)=e^{-\Gamma t}\simeq 1-\Gamma t,\quad \rho_{12}(t)=-\frac{i\Omega t}{2}\sim 0,  \notag
\end{equation}
where a more strict condition $\sqrt{\Gamma^2-16\Omega^2}\gg\Omega$ has been employed in the approximation.
Thus, the relative entropy between the initial state $\rho_0$ and the instantaneous state $\rho$ is
\begin{equation}
D(\rho_0\Vert\rho)=\Gamma t ,\notag
\end{equation}
with $\sigma_{sys}=H(\Gamma t)$, and
\begin{equation}
\sigma_{bath}=\beta (E_2-E_1)\Gamma t=\frac{\Gamma t}{\sin^2\theta_+-\sin^2\theta_-}\ln\frac{\lambda_-}{\lambda_+},   \notag
\end{equation}
where the function $H(s)=-s\ln(s)-(1-s)\ln(1-s)$. Due to $\Gamma\gg\Omega$, we have $\lambda_+\sim 0,\lambda_-\sim 1$ and $\sin(\theta_+)\sim 1, \sin(\theta_-)\sim \Omega/\Gamma_e\sim 0$. So we obtain
\begin{equation}
\sigma_{bath}\gg \Gamma t.\notag
\end{equation}
Similarly, if $\Gamma t\ll 1$ is satisfied, then we can straightforwardly prove that $H(\Gamma t)>\Gamma t$, which means $\sigma_{sys}+\sigma_{bath}>D(\rho_0\Vert\rho)$. So the information-theoretical bound of irreversibility always holds in this case, as shown in the left panel of Fig. \ref{SFig2}. In the small DDR case, dissipation is dominant and thus this relaxation process is nearly classical, with very weak action from the coherence regarding the off-diagonal elements.

\begin{figure}[hbtp]
\centering {\includegraphics[width=9 cm, height=4.5 cm]{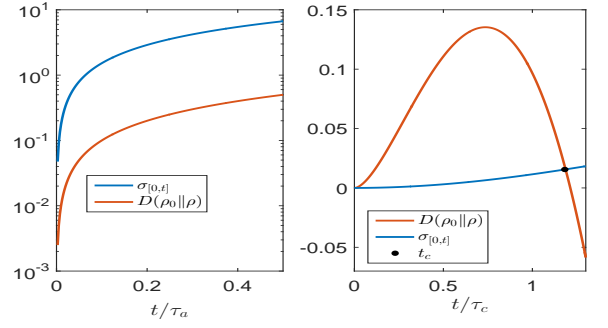}}
\caption{Time evolution of the total entropy production $\sigma_{[0,t]}$ and the relative entropy $D(\rho_0\Vert\rho)$, where the curves are from analytical results.
(Left panel) Small DDR case with $\Omega/\Gamma=0.05$. (Right panel) Ultra-large DDR case with $\Omega/\Gamma=40$ and $t_c=1.15\tau_c$. }
\label{SFig2}
\end{figure}

In contrast to the above, we consider below the ultra-large DDR case, i.e., $\Gamma\ll\Omega$ with $\tau_a\approx 4/3\Gamma$ and $\tau_c\approx\Omega^{-1}$ ($\tau_c\ll\tau_a$). In the case of $t\ll\tau_c$,
\begin{equation}
\rho_{22}(t)=\frac{1}{2}(1+\cos \Omega t),\quad \rho_{12}(t)=-\frac{i}{2}\sin \Omega t,    \notag
\end{equation}
where a more concrete condition $4\Omega\gg 5\Gamma$ has also been used.
Thus the relative entropy between the initial state $\rho_0$ and the instantaneous state $\rho$ is given by
\begin{equation}
D(\rho_0\Vert\rho)\simeq\frac{\Omega^2t^2(\ln 4-1-2\ln \Omega t)}{4},\notag
\end{equation}
with $\sigma_{sys}=0$, and
\begin{equation}
\sigma_{bath}=\frac{\sin^2(\Omega t)}{\sin^2(\Omega t)+(1+\cos\Omega t)^2}\ln\frac{\lambda_-}{\lambda_+}.    \notag
\end{equation}
Due to the condition $\Gamma\ll \Omega$, we have $\lambda_+\sim (1-\Gamma/\Omega)/2$ and $\lambda_-\sim (1+\Gamma/\Omega)/2$ where $\lambda_-/\lambda_+=1+2\Gamma/\Omega$. So we find,
\begin{equation}
\sigma_{bath}= \frac{\sin^2(\Omega t)}{\sin^2(\Omega t)+(1+\cos\Omega t)^2}\frac{2\Gamma}{\Omega}\simeq \frac{\Gamma\Omega}{2}t^2.    \notag
\end{equation}
Thus the condition $\Omega t\ll 1$ yields $D(\rho_0\Vert\rho)>\Omega^2t^2$. Using the fact of $\Omega\gg\Gamma$, we obtain $D(\rho_0\Vert\rho)\gg\sigma_{bath}+\sigma_{sys}$ for $t\ll\tau_c$. Therefore, the generalized information-theoretical bound of irreversibility is violated in the ultra-large DDR case. Moreover, for a time $t_c$ comparable to $\tau_c$, we consider following condition
\begin{equation}
\frac{\Omega^2t_c^2}{4}(\ln 4-1-2\ln \Omega t_c)=\frac{\Gamma\Omega}{2}t_c^2,    \notag
\end{equation}
which violates the information-theoretical bound of irreversibility in the case of $t<t_c$. Solving it, we obtain
\begin{equation}
t_c=\frac{2}{\Omega}e^{-\frac{\Omega+2\Gamma}{2\Omega}}.  \notag
\end{equation}
Thus, in the ultra-large DDR case, the information-theoretical bound of irreversibility can be violated in the first duration ($\le\tau_c$) of the evolution (See right panel of Fig. \ref{SFig2}) since there is a relatively strong coherent oscillation and thus quantum coherent evolution dominates the information interchange between the system and the bath, implying that the relaxation process is quantum mechanical. With the evolution going on, the violation and validity of the bound occur repeatedly.

Considering the analytical results (e.g., $\rho_{12}$ or $\rho_{22}$) as listed above, we may divide our observation into four sections based on following characteristic conditions: $\sqrt{\Gamma^2-16\Omega^2}/\Omega=5$ to separate small DDR from the intermediate; $4\Omega/5\Gamma=5$ to distinguish intermediate DDR from the large; $4\Omega/5\Gamma>20$ to single out the ultra-large DDR. Consequently, we have in the main text four regimes with respect to the DDR values: small (i.e., $\Omega/\Gamma < 0.15$), intermediate (i.e., $0.15 \leq\Omega/\Gamma < 6.25$), large (i.e., $6.25 \leq\Omega/\Gamma < 25$) and ultra-large (i.e., $\Omega/\Gamma\geq 25$).

\section{Effective two-level system from three levels}
Consider the Hamiltonian of a three-level system in the interaction picture,
\begin{equation}
H_s=\frac{1}{2}(\Omega|2\rangle\langle 1|+\tilde{\Omega}|3\rangle\langle 2| +H.C.),   \notag
\end{equation}
where $\Omega$ ($\tilde{\Omega}$) is the Rabi frequency between $|2\rangle$ and $|1\rangle$ ($|3\rangle$). The corresponding Lindblad master equation is given by
\begin{equation}
\dot{\rho}=-i[H,\rho]+\frac{\Gamma_e}{2}(2|1\rangle\langle 3|\rho|3\rangle\langle 1|-|3\rangle\langle 3|\rho-\rho|3\rangle\langle 3|).   \notag
\end{equation}
Defining $\rho_{ij}=\langle i|\rho|j\rangle$, we obtain $\frac{d\rho_{ij}}{dt}=\langle i|\dot{\rho}|j\rangle$. Using this relation, we calculate the dynamic equation of $\rho_{ij}$ as follows,
\begin{eqnarray}
\frac{d}{dt}\rho_{11}&=&-i\frac{\Omega}{2}(\rho_{21}-\rho_{12})+\Gamma_e(1-\rho_{11}-\rho_{22}), \notag \\
\frac{d}{dt}\rho_{22}&=&-i[\frac{\tilde{\Omega}}{2}(\rho_{32}-\rho_{23})+\frac{\Omega}{2}(\rho_{12}-\rho_{21})], \notag \\
\frac{d}{dt}\rho_{12}&=&-i[\frac{\Omega}{2}(\rho_{22}-\rho_{11})-\frac{\tilde{\Omega}}{2}\rho_{13}], \notag \\
\frac{d}{dt}\rho_{21}&=&-i[\frac{\Omega}{2}(\rho_{11}-\rho_{22})+\frac{\tilde{\Omega}}{2}\rho_{31}], \notag \\
\frac{d}{dt}\rho_{13}&=&-i(\frac{\Omega}{2}\rho_{23}-\frac{\tilde{\Omega}}{2}\rho_{12})-\frac{\Gamma_e}{2}\rho_{13}, \notag \\
\frac{d}{dt}\rho_{31}&=&-i(\frac{\tilde{\Omega}}{2}\rho_{21}-\frac{\Omega}{2}\rho_{32})-\frac{\Gamma_e}{2}\rho_{31}, \notag \\
\frac{d}{dt}\rho_{23}&=&-i[\frac{\tilde{\Omega}}{2}(1-\rho_{11}-2\rho_{22})+\frac{\Omega}{2}\rho_{13}]-\frac{\Gamma_e}{2}\rho_{23}, \notag \\
\frac{d}{dt}\rho_{32}&=&-i[\frac{\tilde{\Omega}}{2}(2\rho_{22}+\rho_{11}-1)-\frac{\Omega}{2}\rho_{31}]-\frac{\Gamma_e}{2}\rho_{32}, \notag
\end{eqnarray}
in which we have used the relation $\rho_{11}+\rho_{22}+\rho_{33}=1$. The equations above can be rewritten as
\begin{equation}
\dot{\vec{\rho}}=\mathbf{A}\vec{\rho}-\vec{b},  \notag
\end{equation}
with
\begin{displaymath}
\vec{\rho} =
\left( \begin{array}{ccccccccc}
\rho_{11}  \\
\rho_{22}  \\
\rho_{12} \\
\rho_{21} \\
\rho_{13} \\
\rho_{31} \\
\rho_{23} \\
\rho_{32}
\end{array} \right),\quad
\vec{b} =
\left( \begin{array}{ccccccccc}
-\Gamma_e  \\
0  \\
0 \\
0 \\
0 \\
0 \\
i\frac{\tilde{\Omega}}{2} \\
-i\frac{\tilde{\Omega}}{2}
\end{array} \right),
\end{displaymath}
and
\begin{displaymath}
\mathbf{A} =
\left( \begin{array}{ccccccccc}
-\Gamma_e & -\Gamma_e &  i\frac{\Omega}{2} & -i\frac{\Omega}{2} &0&0&0&0 \\
0 & 0 &  -i\frac{\Omega}{2}&i\frac{\Omega}{2}&0&0&i\frac{\tilde{\Omega}}{2}&-i\frac{\tilde{\Omega}}{2} \\
i\frac{\Omega}{2} & -i\frac{\Omega}{2} &0&0&i\frac{\tilde{\Omega}}{2}&0&0&0\\
-i\frac{\Omega}{2} & i\frac{\Omega}{2} &0&0&0&-i\frac{\tilde{\Omega}}{2}&0&0\\
0 & 0 &i\frac{\tilde{\Omega}}{2}&0&-\frac{\Gamma_e}{2}&0&-i\frac{\Omega}{2}&0\\
0 & 0 &0&-i\frac{\tilde{\Omega}}{2}&0&-\frac{\Gamma_e}{2}&0&i\frac{\Omega}{2}\\
i\frac{\tilde{\Omega}}{2} & 2i\frac{\tilde{\Omega}}{2} &0&0&-i\frac{\Omega}{2}&0&-\frac{\Gamma_e}{2}&0\\
-i\frac{\tilde{\Omega}}{2} & -2i\frac{\tilde{\Omega}}{2} &0&0&0&i\frac{\Omega}{2}&0&-\frac{\Gamma_e}{2}
\end{array} \right).
\end{displaymath}
First, we solve the steady state, i.e., $\dot{\vec{\rho}}=0$, by the equation $\mathbf{A}\vec{\rho}=\vec{b}$. We obtain
\begin{equation}
\rho_{11}=\frac{\Gamma_e^2\Omega^2+\tilde{\Omega}^2(\tilde{\Omega}^2-\Omega^2)+\Omega^4}{2\Gamma_e^2\Omega^2+\tilde{\Omega}^4+2\Omega^4}, ~~ \rho_{22}=\frac{\Gamma_e^2\Omega^2+\Omega^4}{2\Gamma_e^2\Omega^2+\tilde{\Omega}^4+2\Omega^4}, \notag
\end{equation}
\begin{equation}
\rho_{33}=\frac{\Omega^2\tilde{\Omega}^2}{2\Gamma_e^2\Omega^2+\tilde{\Omega}^4+2\Omega^4},~~\rho_{12}=\frac{i\Gamma_e\tilde{\Omega}^2\Omega}{2\Gamma_e^2\Omega^2+
\tilde{\Omega}^4+2\Omega^4}, \notag
\end{equation}
and
\begin{equation}
\rho_{13}=\frac{\tilde{\Omega}\Omega(\Omega^2-\tilde{\Omega}^2)}{2\Gamma_e^2\Omega^2+\tilde{\Omega}^4+2\Omega^4}, ~~ \rho_{23}=\frac{i\Gamma_e\tilde{\Omega}\Omega^2}{2\Gamma_e^2\Omega^2+\tilde{\Omega}^4+2\Omega^4},  \notag
\end{equation}
and $\rho_{21}=\rho_{12}^*$, $\rho_{31}=\rho_{13}^*$ and $\rho_{32}=\rho_{23}^*$.

Assuming the condition $\Gamma_e\gg\tilde{\Omega}\gg\Omega$, we have
\begin{equation}
\rho_{11}=\frac{\Gamma_e^2\Omega^2+\tilde{\Omega}^4}{2\Gamma_e^2\Omega^2+\tilde{\Omega}^4}, ~~\rho_{22}=\frac{\Gamma_e^2\Omega^2}{2\Gamma_e^2\Omega^2+\tilde{\Omega}^4}, \notag
\end{equation}
which lead to $\rho_{22}/\rho_{11}=4\Gamma_e^2\Omega^2/(4\Gamma_e^2\Omega^2+\tilde{\Omega}^4)$, and
\begin{equation}
\rho_{33}=\frac{\Omega^2\tilde{\Omega}^2}{2\Gamma_e^2\Omega^2+\tilde{\Omega}^4}\sim \frac{\Omega}{\Gamma_e}\sim 0 . \notag
\end{equation}
Thus we obtain $\rho_{11}+\rho_{22}=1-O(\Omega/\Gamma_e)\simeq 1$ and $\rho_{11}>\rho_{22}$ no matter how much we adjust $\Omega$ and $\tilde{\Omega}$. Moreover, the off-diagonal terms are given by
\begin{equation}
\rho_{12}=\frac{i\Gamma_e\Omega\tilde{\Omega}^2}{2\Gamma_e^2\Omega^2+\tilde{\Omega}^4}, \notag
\end{equation}
and
\begin{equation}
\rho_{13}=-\frac{2\tilde{\Omega}^3\Omega}{2\Gamma_e^2\Omega^2+\tilde{\Omega}^4}\sim -\frac{\Omega}{\tilde{\Omega}}, ~~ \rho_{23}=\frac{i\Gamma_e\tilde{\Omega}\Omega^2}{2\Gamma_e^2\Omega^2+\tilde{\Omega}^4}\sim \frac{i\Omega}{\tilde{\Omega}}. \notag
\end{equation}
In this context, we find that the predominant terms are $\rho_{11},\rho_{22},\rho_{12}$, which constitutes a two-level system with a small leakage. The steady state of this effective two-level system can be written as
\begin{displaymath}
\rho_s=\frac{1}{2\Gamma_e^2\Omega^2+\tilde{\Omega}^4}
\left( \begin{array}{ccccccccc}
\Gamma_e^2\Omega^2+\tilde{\Omega}^4 & i\Gamma_e\Omega\tilde{\Omega}^2\\
-i\Gamma_e\Omega\tilde{\Omega}^2 & \Gamma_e^2\Omega^2
\end{array} \right).
\end{displaymath}

In what follows, we calculate the effective decay rate in the case of $\Omega=0$. From above equation, if $\Omega=0$, we have
\begin{equation}
\rho_{11}=1,~ \rho_{22}=0,~ \rho_{33}=0,~ \rho_{12}=0,~ \rho_{13}=0,~ \rho_{23}=0, \notag
\end{equation}
corresponding to the situation with a decay from $|2\rangle$ to $|1\rangle$. In this case, the dynamic equation is given by
\begin{eqnarray}
\frac{d}{dt}\rho_{11}&=&\Gamma_e(1-\rho_{11}-\rho_{22}), \quad \frac{d}{dt}\rho_{22}=-i\frac{\tilde{\Omega}}{2}(\rho_{32}-\rho_{23}), \notag \\
\frac{d}{dt}\rho_{12}&=&i\frac{\tilde{\Omega}}{2}\rho_{13}, \quad \frac{d}{dt}\rho_{21}=-i\frac{\tilde{\Omega}}{2}\rho_{31}, \notag \\
\frac{d}{dt}\rho_{13}&=&i\frac{\tilde{\Omega}}{2}\rho_{12}-\frac{\Gamma_e}{2}\rho_{13}, \quad \frac{d}{dt}\rho_{31}=-i\frac{\tilde{\Omega}}{2}\rho_{21}-\frac{\Gamma_e}{2}\rho_{31}, \notag \\
\frac{d}{dt}\rho_{23}&=&-i\frac{\tilde{\Omega}}{2}(1-\rho_{11}-2\rho_{22})-\frac{\Gamma_e}{2}\rho_{23},\notag \\
 \frac{d}{dt}\rho_{32}&=&-i\frac{\tilde{\Omega}}{2}(2\rho_{22}+\rho_{11}-1)-\frac{\Gamma_e}{2}\rho_{32}.\notag
\end{eqnarray}
If the system is initially in $|2\rangle$, we can obtain
\begin{widetext}
\begin{equation}
\rho_{11}=\frac{e^{-\frac{t\Gamma_e}{2}}[8\tilde{\Omega}^2+2e^{\frac{t\Gamma_e}{2}}(\Gamma_e^2-4\tilde{\Omega}^2)+\Gamma_e e^{-\frac{t}{2}\sqrt{\Gamma_e^2-4\tilde{\Omega}^2}}(\sqrt{\Gamma_e^2-4\tilde{\Omega}^2}-\Gamma_e)-\Gamma_e e^{\frac{t}{2}\sqrt{\Gamma_e^2-4\tilde{\Omega}^2}}(\sqrt{\Gamma_e^2-4\tilde{\Omega}^2}+\Gamma_e)]}{2(\Gamma_e^2-4\tilde{\Omega}^2)}. \notag
\end{equation}
\end{widetext}
Using the condition $\Gamma_e\gg \tilde{\Omega}$, we have $\sqrt{\Gamma_e^2-4\tilde{\Omega}^2}=\Gamma_e(1-2\tilde{\Omega}^2/\Gamma_e^2)$ and thus $\rho_{11}$ is reduced to
\begin{equation}
\rho_{11}=1-(1+\frac{3\tilde{\Omega}^2}{\Gamma_e^2})e^{-\frac{\Omega^2}{\Gamma_e}t}+\frac{\tilde{\Omega}^2}{\Gamma_e^2}e^{-\frac{\Gamma_e}{2}t}(4-e^{-\frac{\Gamma_e}{2}t}). \notag
\end{equation}
At $t=0$, $\rho_{11}=0$. For $t\gg 1/\Gamma_e$, the last term turns to be negligible, i.e.,
\begin{equation}
\rho_{11}=1-(1+\frac{3\tilde{\Omega}^2}{\Gamma_e^2})e^{-\frac{\Omega^2}{\Gamma_e}t}\sim 1-e^{-\frac{\Omega^2}{\Gamma_e}t}. \notag
\end{equation}
In the case of $t\ll\Gamma_e/\tilde{\Omega}^2$, $\rho_{11}$ is further reduced to $\rho_{11}\sim \frac{\Omega^2}{\Gamma_e}t$. Similarly,
the solution of $\rho_{33}$ is
\begin{equation}
\rho_{33}=\frac{\tilde{\Omega}^2e^{-\frac{t\Gamma_e}{2}(\Gamma_e+\sqrt{\Gamma_e^2-4\tilde{\Omega}^2})}(e^{\frac{t\Gamma_e}{2}\sqrt{\Gamma_e^2-4\tilde{\Omega}^2}}-1)^2}{(\Gamma_e^2-4\tilde{\Omega}^2)}\simeq \frac{\tilde{\Omega}^2e^{-\frac{\tilde{\Omega}^2}{\Gamma_e}t}}{\Gamma_e^2}, \notag
\end{equation}
and  $\rho_{22}$ is written as
\begin{equation}
\rho_{22}\simeq e^{-\frac{\tilde{\Omega}^2}{\Gamma_e}t}(1-\frac{\tilde{\Omega}^2}{\Gamma_e^2}),  \notag
\end{equation}
for $t\gg 1/\Gamma_e$. Therefore, the effective decay rate from $|2\rangle$ to $|1\rangle$ is given by
\begin{equation}
\Gamma=\frac{\tilde{\Omega}^2}{\Gamma_e}. \notag
\end{equation}
In other words, considering $\tilde{\Omega}^2=\Gamma_e\Gamma$, we have the steady state $\rho_s$ in this case with the same form as the two-level system treated in above section.

\section{Leakage effect in the ultra-large DDR case}
Based on the deductions in above sections, the accurate form of the steady state of the effective two-level system is written as,
\begin{widetext}
\begin{displaymath}
\tilde{\rho}_s=\frac{1}{2\Gamma_e^2\Omega^2+\Gamma_e^2\Gamma^2+2\Omega^4}
\left( \begin{array}{ccccccccc}
\Gamma^2\Gamma_e^2+\Omega^2\Gamma_e(\Gamma_e-\Gamma)+\Omega^4 & i\Gamma_e^2\Gamma\Omega\\
-i\Gamma_e^2\Gamma\Omega & \Gamma_e^2\Omega^2+\Omega^4
\end{array} \right).
\end{displaymath}
\end{widetext}
Under the condition $\Gamma_e\gg\tilde{\Omega}\gg\Omega$, i.e., $\Gamma_e\gg\sqrt{\Gamma\Gamma_e}\gg\Omega$, we obtain
\begin{displaymath}
\tilde{\rho}_s=\rho_s -\xi|1\rangle\langle 1\vert,
\end{displaymath}
with the leakage parameter $\xi=\Omega^2\Gamma/\Gamma_e(2\Omega^2+\Gamma^2)$. In the case of a very small DDR, it is simplified as $\xi=\Omega^2/\Gamma\Gamma_e$, while in the ultra-large DDR case, it reduces to $\xi=\Gamma/2\Gamma_e$ (See left panel of Fig. \ref{SFig3}).

\begin{figure}[hbtp]
\centering {\includegraphics[width=9 cm, height=4.6 cm]{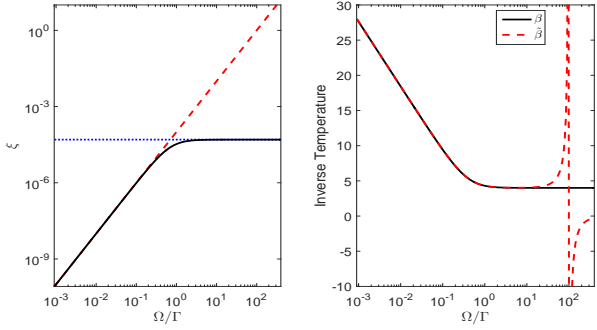}}
\caption{(Left panel) Leakage from an effective two-level system, where blue and red curves represent the ultra-large and small cases, respectively. (Right panel) Inverse temperature regarding a real two-level system (blue curve) and an effective two-level system (red curve), respectively, where the decay rate of the excited state in a three-level system is set as $\Gamma_e=10^4\Gamma$.}
\label{SFig3}
\end{figure}

Writing $\tilde{\rho}_s$ by the eigenspectrum decomposition $\tilde{\rho}_s=\tilde{\lambda}_- |\phi\rangle_{-}\langle\phi|+\tilde{\lambda}_+ |\phi\rangle_{+}\langle\phi|$ with the eigenvalues of the steady state
\begin{equation}
\tilde{\lambda}_{\pm}=\frac{\Gamma^2-r\Omega^2+2\Omega^2\mp\sqrt{4\Gamma^2\Omega^2+(\Gamma^2-r\Omega^2)^2}}{2(\Gamma^2+2\Omega^2)},\notag
\end{equation}
with the decay ratio $r=\Gamma/\Gamma_e$, we may write the corresponding eigenstates as
\begin{equation}
 |\tilde{\phi}\rangle_{+}=(-i\cos\tilde{\theta}_+,\sin\tilde{\theta}_+),\ |\tilde{\phi}\rangle_{-}=(i\cos\tilde{\theta}_-,\sin\tilde{\theta}_-), \notag
\end{equation}
where
\begin{widetext}
\begin{equation}
\tilde{\theta}_{\pm}=\arcsin\frac{2\Gamma\Omega}{\sqrt{2[(\Gamma^2-r\Omega^2)^2+4\Gamma^2\Omega^2\mp(\Gamma^2-r\Omega^2)\sqrt{(\Gamma^2-r\Omega^2)^2
+4\Gamma^2\Omega^2}]}}  \notag
\end{equation}
\end{widetext}
satisfy $\tilde{\theta}_+>\tilde{\theta}_-$ and $\tilde{\theta}_-+\tilde{\theta}_+=\pi/2$. The inverse temperature is given by
\begin{widetext}
\begin{equation}
\tilde{\beta}=\frac{2}{E_2-E_1}\frac{\sqrt{(\Gamma^2-r\Omega^2)^2+4\Gamma^2\Omega^2}}{\Gamma^2-r\Omega^2}\ln\frac{\Gamma^2-r\Omega^2+2\Omega^2+
\sqrt{4\Gamma^2\Omega^2+(\Gamma^2-r\Omega^2)^2}}{2(1-r)\Omega^2}. \notag
\end{equation}
\end{widetext}
In the case of $r=0$, we have $\tilde{\beta}=\beta$. Moreover, in the limit of $\Gamma\gg\Omega$, the above equation reduces to
\begin{equation}
\tilde{\beta}=\frac{4}{ (E_2-E_1)}\ln\frac{\Gamma}{\Omega}, \notag
\end{equation}
which is the same as that for a real two-level system, as shown in right panel of Fig. \ref{SFig3}. However, in the $\Gamma\ll\Omega$, we find
\begin{equation}
\tilde{\beta}=\frac{4}{(E_2-E_1)} \frac{\Gamma^2}{\Gamma^2-r\Omega^2}, \notag
\end{equation}
which is different from the case of a real two-level system. This difference is vanishing if $r\Omega^2\ll\Gamma^2$, implying the condition $\Omega^2\ll\Gamma\Gamma_e$.  As a result, in the case of ultra-large DDR, we have to satisfy the condition $\Gamma\ll\Omega\ll\sqrt{\Gamma\Gamma_e}$, i.e., $\Gamma_e/\Omega\sim (\Omega/\Gamma)^3$. Unfortunately, this is a very challenging condition for an effective two-level system since violating the bound of irreversibility requires $\Omega/\Gamma\ge 36$, indicating that $\Omega\le\Gamma_e/36^3$ and $\Gamma\le\Gamma_e/36^4$. Since $\Gamma_{e}$ is generally of the order of tens of MHz in atomic systems, both $\Omega$ and $\Gamma$ required would be less than 500 Hz, which means impossibility to accomplish qualified optic measurements.

The above discussions are based on the assumption that the system is initialized from $|2\rangle$. In fact, as a quantum system, the state can be initialized in any superposition or mixed state, rather than simply in a well-polarized state. If the system is initially prepared in a state with coherence, the violation of the bound can be observed in the regimes of the small, intermediate and large DDR, as investigated in section IV (B) of the main text.

On the other hand, if we only focus on the condition for a ultra-large DDR, we may consider a realistic two-level system with a fixed decay rate. As such, Rydberg atoms, with effective Rabi frequency possibly larger than the decay rate by more than 50 times, might be available to witness the bound violation we predicted. For example, in a recent publication Phys. Rev. Lett. $\textbf{121}$, 123603 (2018) using $^{87}$Rb atoms, the Rydberg state $|70S, J=1/2; m_{J}=-1/2\rangle$ is of the lifetime 27 $\mu$s, an effective Rabi frequency has reached 2$\pi\times$ 2 MHz, implying $\Omega/\Gamma>$ 54. A larger DDR has also been achieved in a recent neutral atom experiment [Science $\textbf{365}$, 570 (2019)] based on a two-photon process, in which the Rabi frequency can be varied from 0 to 2$\pi\times$~5 MHz and implies that any value of $\Omega/\Gamma$ in the interval (0, 135] is feasible. Nevertheless, to witness the bound violation as we predicted, we consider that more elaborate investigation is required for the experimental operation and measurement, such as the influence of the two-photon process involved, the coherence time remained, and qualified steady state to be reached. For example, laser scattering from the intermediate state of the two-photon process affects our observation. To reduce the laser scattering from the intermediate state, one can employ higher laser powers and further enlarge the detuning from the intermediate state. On the other hand, one can employ the single-photon process to address this issue since the intermediate state is not employed, as shown in a recent experiment [Nature Physics \textbf{12}, 71 (2016)]. The Rabi frequency can reach $\Omega/2\pi=4.3$~MHz, and the lifetime of single Rydberg state in this experiment is measured as $40~\mu$s. The resulted  value of $\Omega/\Gamma$ is about 172.

\begin{figure}[hbtp]
\centering {\includegraphics[width=9.0 cm, height=5.2 cm]{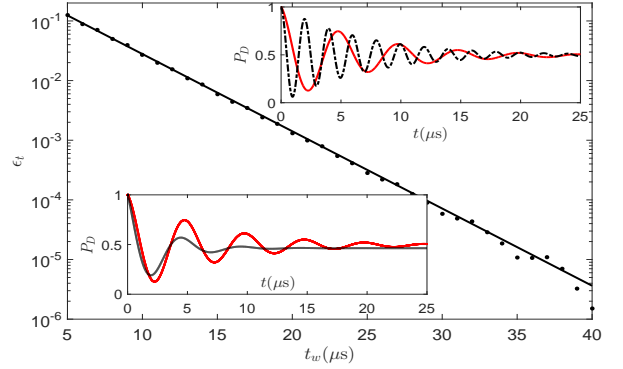}}
\caption{Deviation from the steady state with respect to the waiting time $t_{w}$ in the case of the effective decay of $|2\rangle$. Inset: (top) Time evolution of $P_{D}$ for a waiting time $t_{w}=25~\mu$s with $\Gamma=200$ kHz, where the two curves are with $\Omega/2\pi=200$ (red curve) and $500$ kHz (black curve); (bottom) Time evolution of $P_{D}$ for a waiting time $t_{w}=25 \mu$s with $\Omega/2\pi=200$ kHz, where the two curves are with $\Gamma=200$ kHz (red curve) and $500$ kHz (black curve). }
\label{SFig4}
\end{figure}

\section{Realistic consideration of ultra-large DDR case and the steady state}
The divergence of the inverse temperature $\tilde{\beta}$ exemplified in the right panel of Fig. \ref{SFig3} occurs in the case of $\Gamma_{e}=10^{4}\Gamma$ and $\Omega> 20\Gamma$, which implies that the condition $\Gamma_e/\Omega\sim (\Omega/\Gamma)^3$ is no longer satisfied, i.e., invalidity of an effective two-level system. As discussed in last section, in the case of the initial state in $|2\rangle$, we require $\Gamma\le\Gamma_e/36^4$ and $\Omega\le\Gamma_e/36^3$ in the ultra-large DDR case, which ensures an effective two-level system even for $\Omega/\Gamma\ge 36$ in order to witness the violation of the irreversibility bound.

Moreover, theoretically, the steady state is achieved at $t\rightarrow\infty$. This is, however, impossible for practical operations experimentally. To assess how well for a long time waiting approaching the steady state, we consider, for a particular waiting time, the deviation from the steady state by defining $\epsilon_t=D(\rho_{t}\Vert\rho_s)$. As shown in the top inset of Fig. \ref{SFig4}, smaller Rabi frequency corresponds to less oscillation, but has no change for the waiting time. In contrast, in the bottom inset, we find that larger decay rate produces smaller deviation with an approximate relation $\epsilon_t\sim e^{-\Gamma t_w}$. Therefore, for a given deviation, we obtain a relation for the waiting time $t_w\sim -\Gamma^{-1}\ln\epsilon_t$. For example, for $\epsilon_t=10^{-3}$, the waiting time should be $t_w\sim 7\Gamma^{-1}$.
\begin{figure}[hbtp]
\centering {\includegraphics[width=9.0 cm, height=5.6 cm]{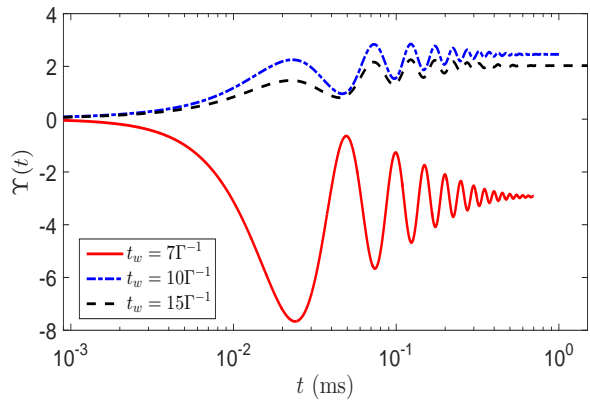}}
\caption{Time evolution of the balance parameter $\Upsilon(t)$ regarding different time scales, where the effective decay rate and the coupling strength are set as $\Gamma=10$ kHz and $\Omega/2\pi=20$ kHz. The decay rate from the excited state is $\Gamma_e/2\pi=23.1$ MHz. The parameters produce $\tilde{\Omega}/2\pi=192$ kHz, satisfying the condition $\Gamma_e\gg\tilde{\Omega}\gg\Omega\gg\Gamma$.}
\label{SFig5}
\end{figure}

In this context, we analyze here another reason regarding the waiting time for the impossibility to realistically witness the violation of Eq. (1) in the ultra-large DDR case. Assuming we can achieve the ultra-large DDR case, we find large inaccuracy in measuring $\Upsilon(t)$ experimentally, which originates from the steady state achieved. In terms of our numerical results in Fig. \ref{SFig5}, we obtain completely different results of $\Upsilon(t)$ for different waiting times. For example, for $t_w=$7, 10, 15$\Gamma^{-1}$, the inverse temperature $\tilde{\beta}=$-5.82, 4.92, 4.06, while the deviations from the steady are $\epsilon_t=1.4\times 10^{-5},3.2\times 10^{-7}$ and $4\times 10^{-8}$, respectively. Therefore, witnessing the violation of the bound in this case requires a much long waiting time, which is very challenging.

\section{Some points for Experimental operations}

All the experimental evolutions are carried out in a simplest system of single spin consisting of an ultracold $^{40}$Ca$^{+}$ ion, confined in a linear Paul Trap. The 397-nm laser is used for three dimensional Doppler cooling, whereas lasers, at wavelengths 866 nm in z- and y-directions and 854 nm in z-direction, are purposed as repumping and state quenching light sources. The 866-nm laser is always on throughout our experiments. The 729-nm laser, in x-direction, performs state manipulation and experimental evolution in addition to sideband cooling so as to bring the trapped ion down to vibrational ground state. The 729-nm laser is an ultra-stable narrow linewidth Ti:Sapphire laser, corresponding to linewidth (FWHM) of 7 Hz, as measured via the heterodyne beat note method with respect to another laser system. It is locked to a high-finesse ultra-low expansion cavity via PDH locking method which could reach a long-term drift less than 0.06 Hz/s.

In order to provide laser cooling, initialize quantum state and perform experimental evolutions and final state detection, we need a well-ordered laser pulse train with polarization, phase, frequency and amplitude control. Therefore, the lasers are controlled via the acousto-optic modulators (AOMs) by passing all the lasers through the AOMs before irradiating the ion. The operational systematic RF signals applied to all the AOMs are being supplied by the direct digital synthesizer (DDS) which is controlled by a field programmable gate array in addition to a TTL architecture intended to provide control signals to the RF switches. The DDS has the ability to provide the AOM signals with real time phase, frequency and amplitude control of all the lasers during the consecutive experimental progresses. A typical experimental sequence involves more than 300 optical pulses within a time slot of about 40 ms.

\end{document}